\documentclass[aps,pra,superscriptaddress,showpacs,showkeys,a4paper,10pt,notitlepage,twocolumn]{revtex4-1}
\usepackage[utf8]{inputenc}
\usepackage[english]{babel}
\usepackage{amsmath}
\usepackage{amssymb}
\usepackage{amsfonts}
\usepackage{graphicx}
\usepackage{braket}
\usepackage{cancel}
\usepackage{upgreek}
\usepackage{mathtools}
\usepackage[T2A]{fontenc}

\usepackage{xcolor}
\definecolor{link_blue}{RGB}{52,46,157}

\usepackage{longtable}
\usepackage[colorlinks,citecolor=link_blue,linkcolor=link_blue,urlcolor=link_blue]{hyperref}
\usepackage[tiny,center,uppercase]{titlesec}

\setcitestyle{super}

\renewcommand{\vec}{\boldsymbol}

\begin{document}

\title{Eigenstates of two-level systems in a single-mode quantum field: from quantum Rabi model to $N$-atom Dicke model}

\author{A.\ U.\ Leonau}
\email[Corresponding author: ]{a.u.leonau@gmail.com}
\affiliation{Belarusian State University, 4 Nezavisimosty Ave.,
  220030, Minsk, Belarus}

\author{N.\ Q.\ San}
\affiliation{Belarusian State University, 4 Nezavisimosty Ave.,
  220030, Minsk, Belarus}

\author{A.\ P.\ Ulyanenkov}
\affiliation{Atomicus GmbH, Schoemperlen Str. 12a, 76185 Karlsruhe, Germany}

\author{O.\ D.\ Skoromnik}
\affiliation{ }

\author{I.\ D.\ Feranchuk}
\affiliation{Belarusian State University, 4 Nezavisimosty Ave.,
  220030, Minsk, Belarus}

\begin{large}

\begin{abstract}
In the present paper we show that the Hamiltonian describing the resonant interaction of $N$ two-level systems with a single-mode electromagnetic quantum field in the Coulomb gauge can be diagonalized with a high degree of accuracy using a simple basis set of states. This allows one to find an analytical approximation for the eigenvectors and eigenvalues of the system, which interpolates the numerical solution in a broad range of the coupling constant values. In addition, the introduced basis states provide a regular way of calculating the corrections and estimating the convergence to the exact numerical solution. The obtained results are valid for both  quantum Rabi model ($N = 1$) and the Dicke model for $N \geq 2$ atoms. 
\end{abstract}

\maketitle



\section{Introduction}
\label{sec:intro}

The quantum Rabi model (QRM) is one of the fundamental models in quantum physics describing the interaction of electromagnetic radiation and matter \cite{PhysRev.49.324,PhysRev.51.652,Braak_2016}. This model incorporates a two-level system (TLS), which interacts with a single mode quantum field in a cavity. QRM proved to be very effective in predicting and describing various physical phenomena \cite{Walther_2006,RevModPhys.73.565,holstein_studies_1959,pedernales_quantum_2015}, and some of its applications are currently widely used \cite{feranchuk_physical_2016,PhysRevA.95.063849,RevModPhys.91.025005,PhysRevA.97.013851}. A generalization of this model for the $N$-TLS case was introduced by Dicke \cite{PhysRev.93.99} for the description of light-matter interaction \cite{dm_revisited_2011,PhysRevA.97.043858,Zhiqiang:17,Scattering_2020}.

To effectively utilize the QRM and Dicke models (DM) in various applications, it is necessary to calculate the eigenvectors and eigenvalues of the associated Hamiltonians in the wide range of coupling constant values, including the ultrastrong and deep strong coupling regimes between the TLS and the field \cite{frisk_kockum_ultrastrong_2019,RevModPhys.91.025005,feranchuk_strong_2011} where the perturbation theory as well as the rotating wave approximation become inapplicable. It was shown recently \cite{PhysRevLett.107.100401} that QRM is an exactly integrable system, and the problem of finding its stationary states for arbitrary values of the coupling constant can be expressed in terms of polynomial recurrence relations. This result is fundamentally important, however, the eigenvectors and eigenvalues are not represented in a closed analytical form, which makes it rather difficult to use them for applications.

Although it is possible to find the numerical solution of the Schrödinger equation with the  QRM Hamiltonian, a large number of works is devoted to the development of the approximate analytical representation of eigenvalues and eigenfunctions in a broad range of  the system parameters \cite{feranchuk_two-level_1996,PhysRevLett.99.173601,PhysRevLett.99.259901,PhysRevA.87.033827,PhysRevA.104.033712}. Analytical solutions can be useful for the classification of the quantum states and  describing the time evolution and thermodynamics, when summation over the entire spectrum of the system  is needed. However, there exists a significant drawback of the currently available approximations, which consists in using the complicated basis sets of functions. As a result, it becomes rather difficult to carry out the analytical assessment of the accuracy  and investigate their convergence to the exact result.

One more important problem of using the QRM and DM and, consequently, the explicit form of the analytical approximations for their eigenvalue problem is related to the validity of the two-level approximation for the Hamiltonian of the matter part.  This problem was considered in the book \cite{scully_quantum_1997} within the framework of perturbation theory and recently was analyzed in the work \cite{di_stefano_resolution_2019}. In particular, it was shown that the finite-level truncation of the matter system leads to the violation of the gauge invariance, and the validity of the two-level approximation becomes gauge-dependent. However, it is possible to restore the gauge invariance by introducing the unitary operator, which transforms the QRM and DM Hamiltonians from the standard dipole gauge to the Coulomb gauge. This takes into account all orders of the vector potential $\vec{A}$ in the interaction Hamiltonian. Disregarding the gauge invariance leads to some contradictory predictions and paradoxes, especially in the cases when the light-matter interaction becomes very strong \cite{PhysRevA.98.053819,stokes_gauge_2019}.

The Hamiltonians obtained in the Coulomb gauge contain exponential operators and have a rather complicated form \cite{di_stefano_resolution_2019}, so that the numerical calculation of the spectrum of their eigenstates requires the use of high dimensional matrices. At the first glance, it seems that this also obstructs the development of analytical approximations. However, we show that it is possible to find a very simple set of basis vectors, which allows one to diagonalize the QRM and DM Hamiltonians in the Coulomb gauge with a sufficiently high accuracy. As a result, we introduce the analytical approximation for the energy levels of the system and its eigenstates within the entire range of variation of the system's parameters. The developed approach is based on the utilization of the operator method (OM) of solving the Schrödinger equation \cite{feranchuk_non-perturbative_2015}, which allows one to find not only the zeroth-order approximation for the energy levels, but also assessing the accuracy of the obtained results and investigating their convergence to the exact solution. It is shown that analytical approximations can be used for practical applications only for the systems with 2- and 3-TLS. For the systems that contain a larger number of TLS it is more efficient to diagonalize  the DM Hamiltonians numerically in the matrix representation, which can be obtained by means of using the simple basis states.

The paper is organized as follows. In Sec.~\ref{sec:2}, the analytical approximation for the eigenvalues of QRM in the Coulomb gauge is deduced. The basis set of states is also constructed, which makes it possible to calculate the successive corrections to the zeroth-order approximation. The analytical results are compared with the numerical solutions, and the rapid convergence of the successive approximations is demonstrated. In Sec.~\ref{sec:3},  the proposed method is applied to the 2- and 3-TLS DM in order to analyze its features. The zeroth-order approximation for the energy levels is developed and its comparison with the numerical results is carried out. In Sec.~\ref{sec:4} we consider the features of calculating the spectrum of the DM with a larger number ($N \gg 1$) of TLS. It is shown that in this case the analytical approximations are cumbersome that makes their practical application ineffective. However, using the introduced basis set allows us to construct the matrix representation of the DM Hamiltonian and perform its numerical diagonalization effectively.

\section{Eigenstates of QRM in the Coulomb gauge}
\label{sec:2}

It is well known that the interaction of a single mode quantum field with an atomic system has a resonant character when the field frequency is close to one of the transition frequencies between some atomic levels. Isolating this pair of states is fundamental to QRM, in which the dipole approximation is employed, when the interaction of the atom with the field is determined by the operators $ (\vec{d} \cdot \vec{E})$, where $\vec{d}$ is the transition dipole moment, and $\vec{E}$ is the electric field. These assumptions lead to the so-called dipole gauge of the QRM, which corresponds to the following Hamiltonian (the natural units $\hbar = c = 1$ are used):
\begin{eqnarray}
\label{1}
	\hat H = \hat{a}^+\hat{a} + \frac{\Delta}{2}\hat{\sigma}_z + f (\hat{a} + \hat{a}^+)\hat{\sigma}_x + f^2,
\end{eqnarray}
where $\hat{a}$, $\hat{a}^+$ are the annihilation and creation operators of the resonant single-mode quantum field with frequency $\omega$, which is chosen as the unit of  energy (that is $\omega = 1 $); $\Delta$ is the resonant transition frequency; $f \sim |\vec d|$ is the dimensionless TLS-field coupling constant; $\hat{\sigma}_{x,z}$ are  Pauli matrices. It should be noted that operator (\ref{1}) differs from the definition used in \cite{di_stefano_resolution_2019} by a canonical transformation $\hat{a} \leftrightarrows -i \hat{a}$, $\hat{a}^+ \leftrightarrows i \hat{a}^+$. In addition, a constant shift $f^2$ is added to the operator (\ref {1}), which, according to \cite{di_stefano_resolution_2019}, must be taken into account when truncating the Hilbert space of the initial atomic system to TLS model.

The Coulomb gauge of QRM arises when the momentum operator in the kinetic energy of the atom is replaced by $ \vec p \rightarrow (\vec p - e \vec A) $, where $e$ is the electron charge, and $\vec A$ is the vector potential of the field. In this case the operator of potential energy  $V(\vec r)$ remains unchanged. After such a replacement, the interaction of the atom with the field is $ \sim (\vec p \cdot \vec A) $, and the so-called diamagnetic term $\sim A^2 $ arises in the Hamiltonian. As a result, this leads to the modification of the QRM Hamiltonian. However, since the electromagnetic interaction is invariant under the gauge transformations, the Hamiltonians both in the dipole and Coulomb gauges should be equivalent. This was shown in \cite{scully_quantum_1997} within the framework of the perturbation theory over the field, but a rigorous proof of the equivalence of both representations was obtained in the recent work \cite{di_stefano_resolution_2019}. The main idea of this proof is related to the fact that the truncation of the Hilbert space of the atomic many-level system to just only two states leads to the nonlocality of the potential $ V (\vec r) $. As a result, it does not commute with the gauge transformation operator, which is determined by the following unitary operator in the framework of the dipole approximation
\begin{eqnarray}
\hat{U} = e^{ f \hat{\sigma}_x (\hat{a}^+ - \hat{a})}.
\end{eqnarray}

Applying this unitary transformation to the operator (\ref{1}), one can derive \cite{di_stefano_resolution_2019} the following expression for the Hamiltonian of QRM in the Coulomb gauge:
\begin{multline}
\label{3}
	\hat H_C = \hat{a}^+ \hat{a} + \frac{\Delta}{2}\Bigl\{\hat{\sigma}_z \cosh[2f(\hat{a} - \hat{a}^+)] \\ 
		- i \hat{\sigma}_y \sinh[2f(\hat{a}-\hat{a}^+)]\Bigr\}.
\end{multline}

Both Hamiltonians (\ref{1}) and (\ref{3}) commute with the combined parity operator
\begin{eqnarray}
\label{4}
\hat P = \hat{\sigma}_z e^{i \pi \hat{a}^+ \hat{a}},
\end{eqnarray}
which should be taken into account when constructing various approximations.

As it was mentioned above, various methods have been proposed to obtain the approximate analytical solution of the Schrödinger equation with Hamiltonian (\ref{1}) in a wide range of the coupling constant values \cite{feranchuk_two-level_1996,PhysRevLett.99.173601,PhysRevLett.99.259901,PhysRevA.87.033827,PhysRevA.104.033712}. All of them are based on complex variational state vectors and transformations, including a number of additional parameters. As an example, the recent work \cite{PhysRevA.104.033712} can be considered. It is also essential that the complex form of the basis set vectors in the zeroth-order approximation does not allow one a straightforward calculation of the corrections to these solutions.

In contrast to this, it seems surprising that Hamiltonian (\ref{3}) which has a more sophisticated form but is equivalent to (\ref{1}), can be diagonalized with a high degree of accuracy using the simple basis set
\begin{eqnarray}
\label{5}
	\ket{\psi_{ns}} = \ket{n} \chi_s; \quad  n = 0,1\ldots;  \ s = \uparrow, \downarrow,
\end{eqnarray}
where $\ket{n}$ are the Fock states of the field; $\chi_s$ are the eigenvectors of the operator $\sigma_z$.  In this basis the spin and the field variables are separated, and its vectors also form the eigenstates of the combined parity operator.

For further calculations we need the matrix elements of Hamiltonian (\ref{3}) in the basis (\ref{5}), which can be derived in the form as follows:
\begin{align}
\label{qrm-matr}
	H_{kn} = n \delta_{kn} \text{I}_2 &+ \frac{\Delta}{2} S_{kn} \biggl[ \frac{(-1)^n + (-1)^k}{2} \sigma_z \nonumber \\ 
	& - i \frac{(-1)^n - (-1)^k}{2} \sigma_y \Bigr], \\
	S_{kn}(f) = (-1)^n & \sqrt{\frac{n!}{k!}} (2f)^{k-n}L_n^{k-n}(4f^2) e^{-2f^2}; \nonumber \\
	 k \geq n; &\ S_{kn}= S_{nk}, \nonumber
\end{align}
where $L_n^k (x)$ are the generalized Laguerre polynomials; $\text{I}_2$ is a unit $2 \times 2$ matrix; $\sigma_y$ and $\sigma_z$ are the $2 \times 2$ Pauli matrices defined as:
\begin{equation}
	\sigma_y = 
	\begin{pmatrix}
	0 & -i \\
	i & 0 
	\end{pmatrix}, \quad
	\sigma_z = 
	\begin{pmatrix}
	1 & 0 \\
	0 & -1 
	\end{pmatrix}.
\end{equation}

\begin{figure*}[t]
  \includegraphics[width=.4\linewidth]{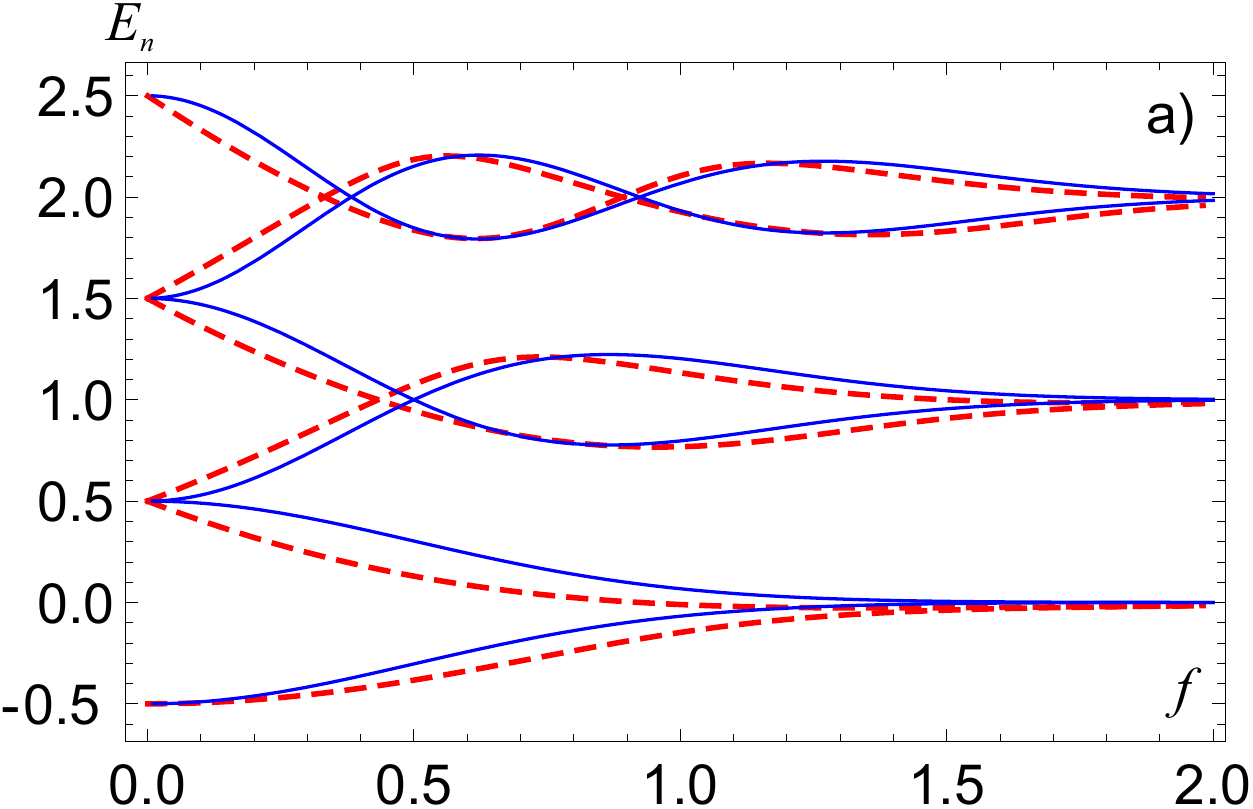} \quad
  \includegraphics[width=.4\linewidth]{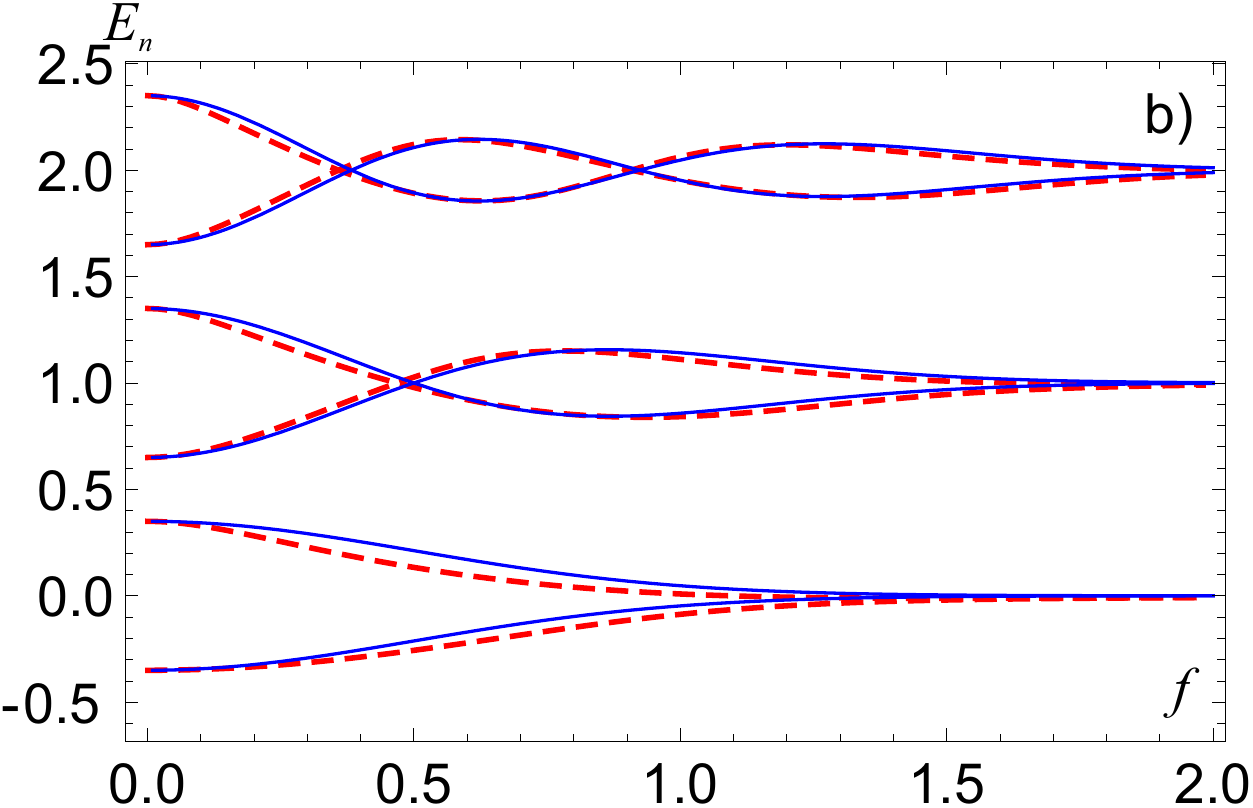} \break\break
  \includegraphics[width=.4\linewidth]{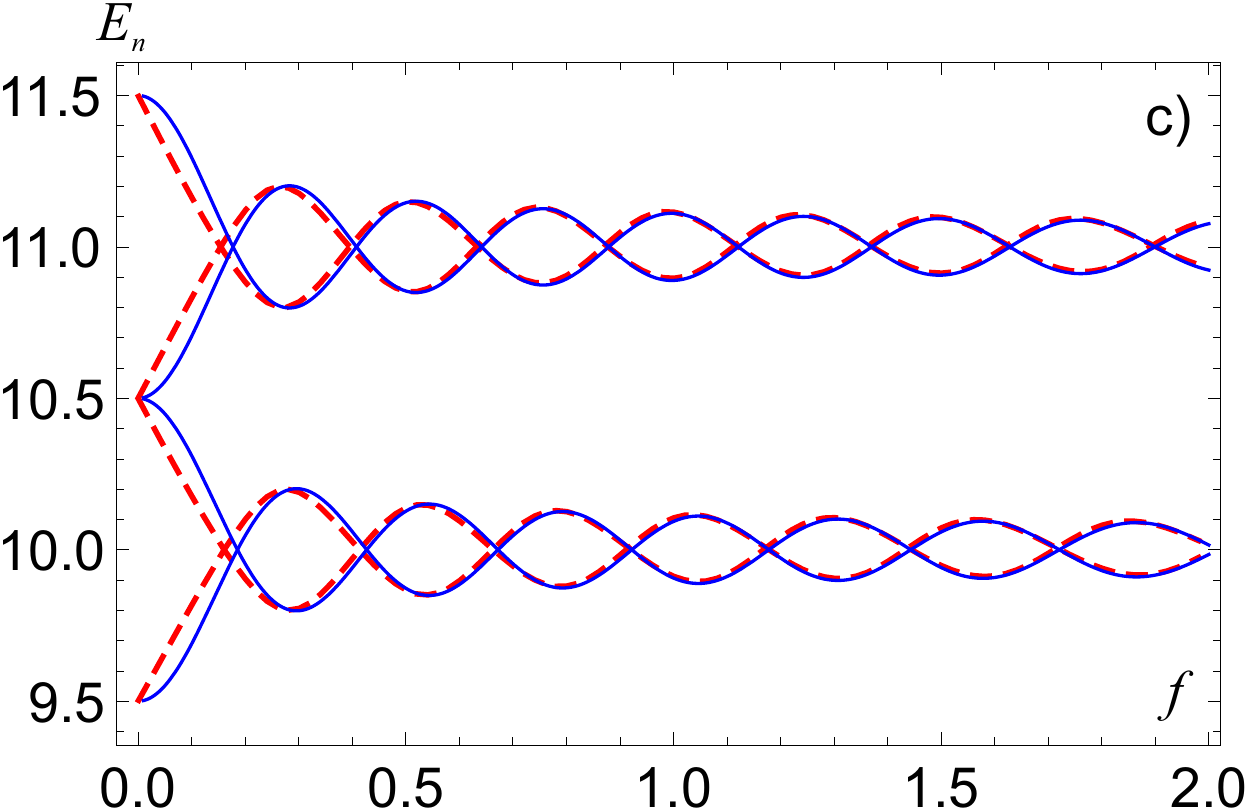} \quad
  \includegraphics[width=.4\linewidth]{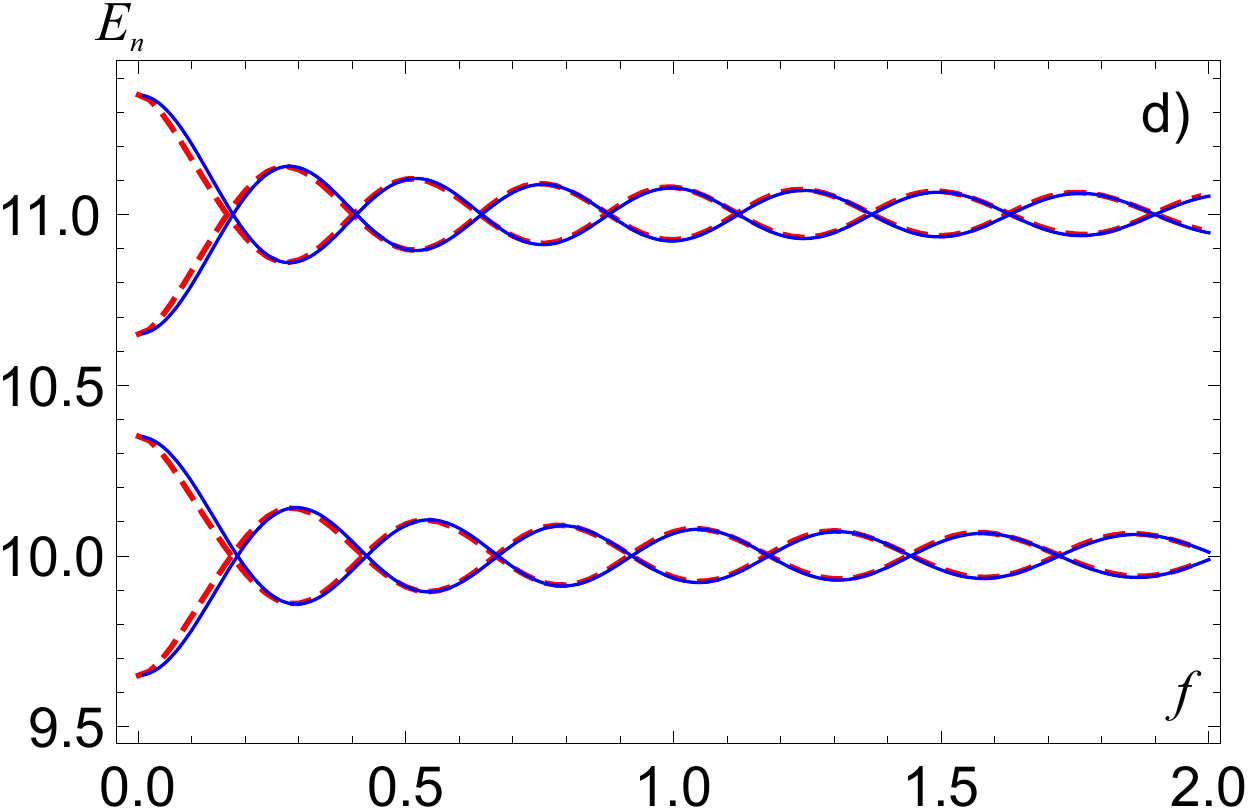}
  \caption{(Color online) Energy levels of QRM as a function of the dimensionless coupling constant $f$ and parameter $\Delta$: (left column) $\Delta = 1.0$, (right column) $\Delta = 0.7$. Low-lying states are shown in (a) and (b), whereas some highly-excited states are presented in (c) and (d). Dashed lines correspond to the numerical solution, solid lines represent the zeroth-order approximation of OM (\ref{7}).} 
    \label{fig:1}
\end{figure*}

A simple basis set (\ref{5}) allows one to use the OM \cite{feranchuk_non-perturbative_2015}, in which the solution of the Schrödinger equation
\begin{eqnarray}
\label{6}
	\hat{H}_C \ket{\Psi_{\nu}} = E_{\nu} \ket{\Psi_{\nu} },
\end{eqnarray}
can be calculated by using the iteration scheme, which ensures the convergence of the successive approximations (here and below the composite index $ \nu = (n, s) $ is used for shortness).

Let us introduce the formula for the zeroth-order approximation of OM for the energy levels, which is given by the diagonal elements of the Hamiltonian matrix (\ref{qrm-matr}) in the considered basis set:
\begin{align}
\label{7}
	E_\nu^{(0)} = H_{\nu \nu} \, \rightarrow \, E_{ns}^{(0)} &= \braket{\psi_{ns}|\hat{H}_C|\psi_{ns}} \nonumber \\
	 &= n + s \frac{\Delta}{2} S_{nn} (f),
\end{align}
where the numerical values $s = +1,-1$ match the spin indices $\uparrow$, $\downarrow$, respectively. 

In Fig.~\ref{fig:1} we plot the results of calculation by using the formula (\ref{7}) for various values of the system's parameters. Comparing the results obtained within the zeroth-order approximation with the results of the numerical calculation, one can conclude that the simple formula (\ref{7}) describes all features of the energy spectrum of the system within the entire range of the dimensionless coupling constant values, and the accuracy improves even further for highly excited states and varying the detuning parameter from the resonant value $(\Delta = 1)$.

At the same time, there is a noticeable deviation from the numerical solution for the resonant case in the region where the dimensionless coupling constant $f$ is small. However, it is still possible to improve the accuracy of the zeroth-order approximation by slightly changing the basis set. For this purpose, one should take into account the degeneracy of the eigenstates having the same parity in the weak coupling region. Similarly to the rotating wave approximation \cite{jaynes_comparison_1963},  it is necessary to use the correct linear combination of the degenerate states, which can be written explicitly as:
\begin{eqnarray}
\label{9}
	\ket{\widetilde{\psi}_n} =  A_n \ket{n} \chi_{\uparrow} + B_n \ket{n + 1} \chi_{\downarrow}.
\end{eqnarray}

\begin{figure*}[t]
  \includegraphics[width=.4\linewidth]{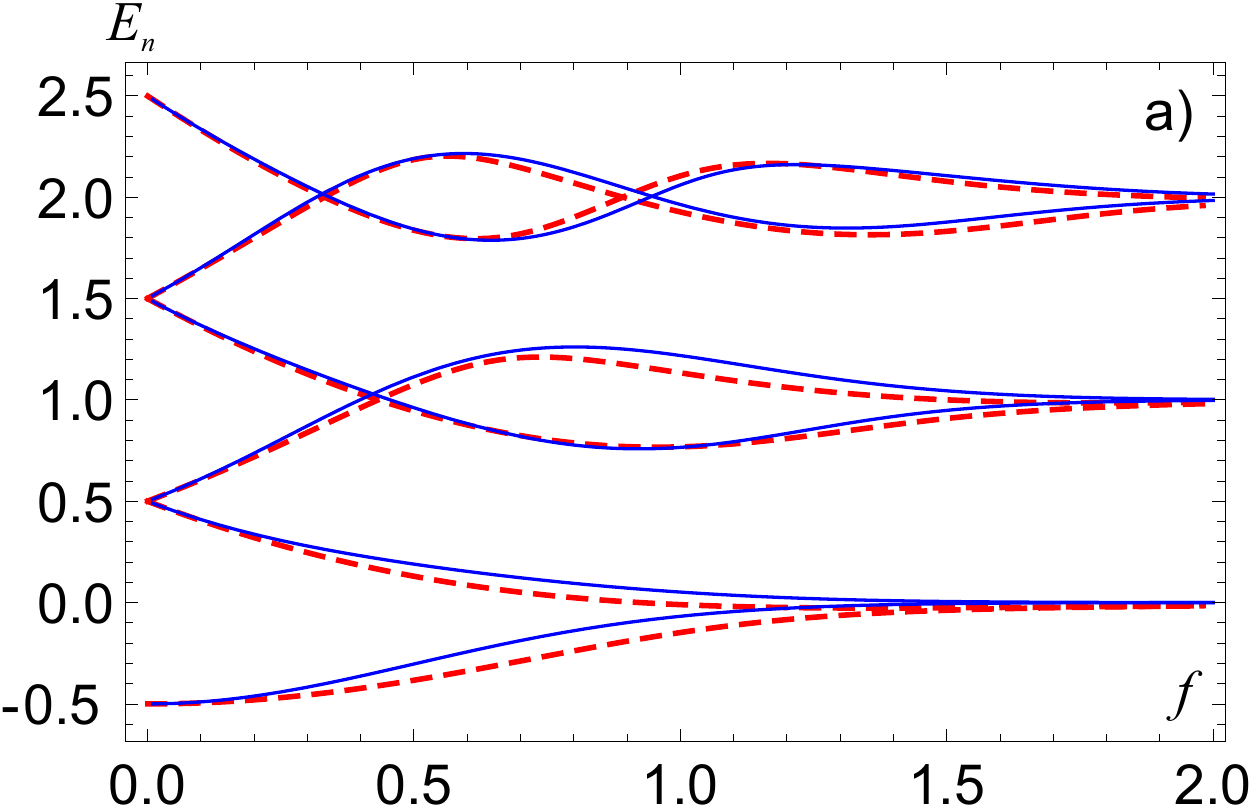} \quad
  \includegraphics[width=.4\linewidth]{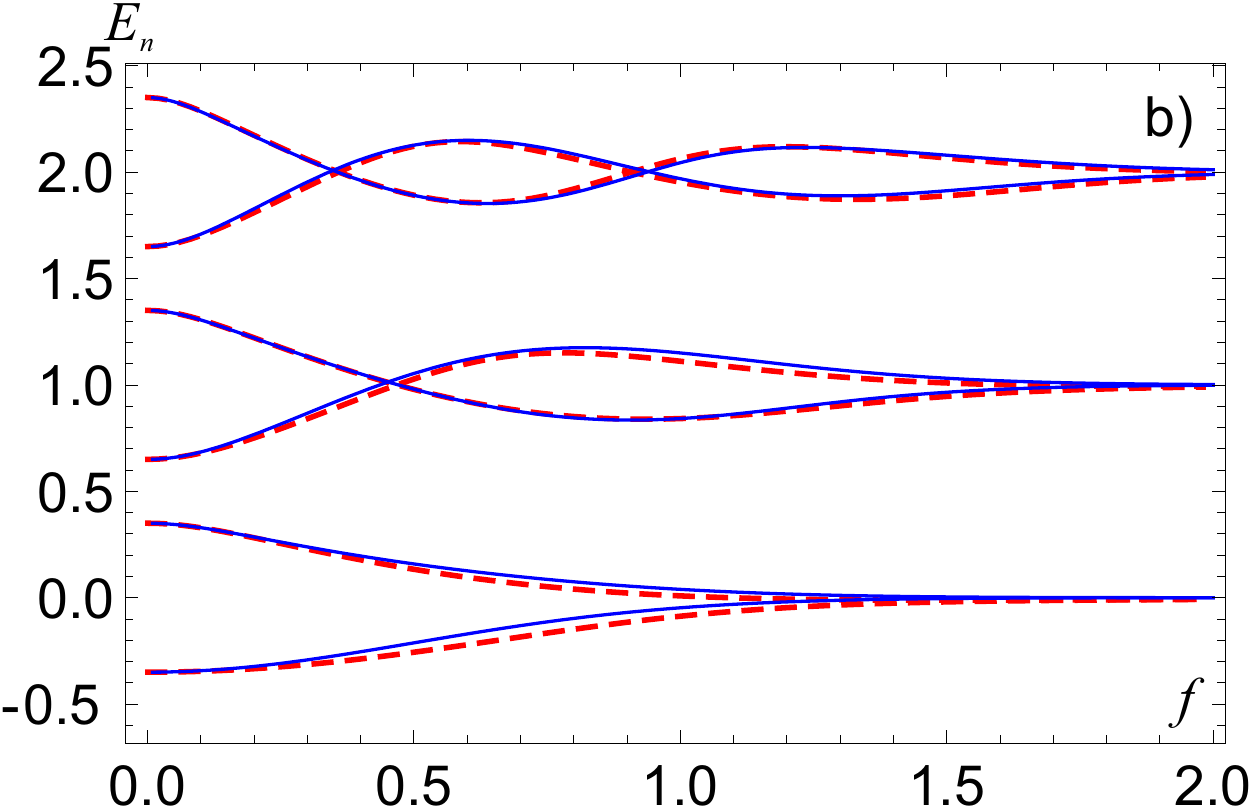} \break\break
  \includegraphics[width=.4\linewidth]{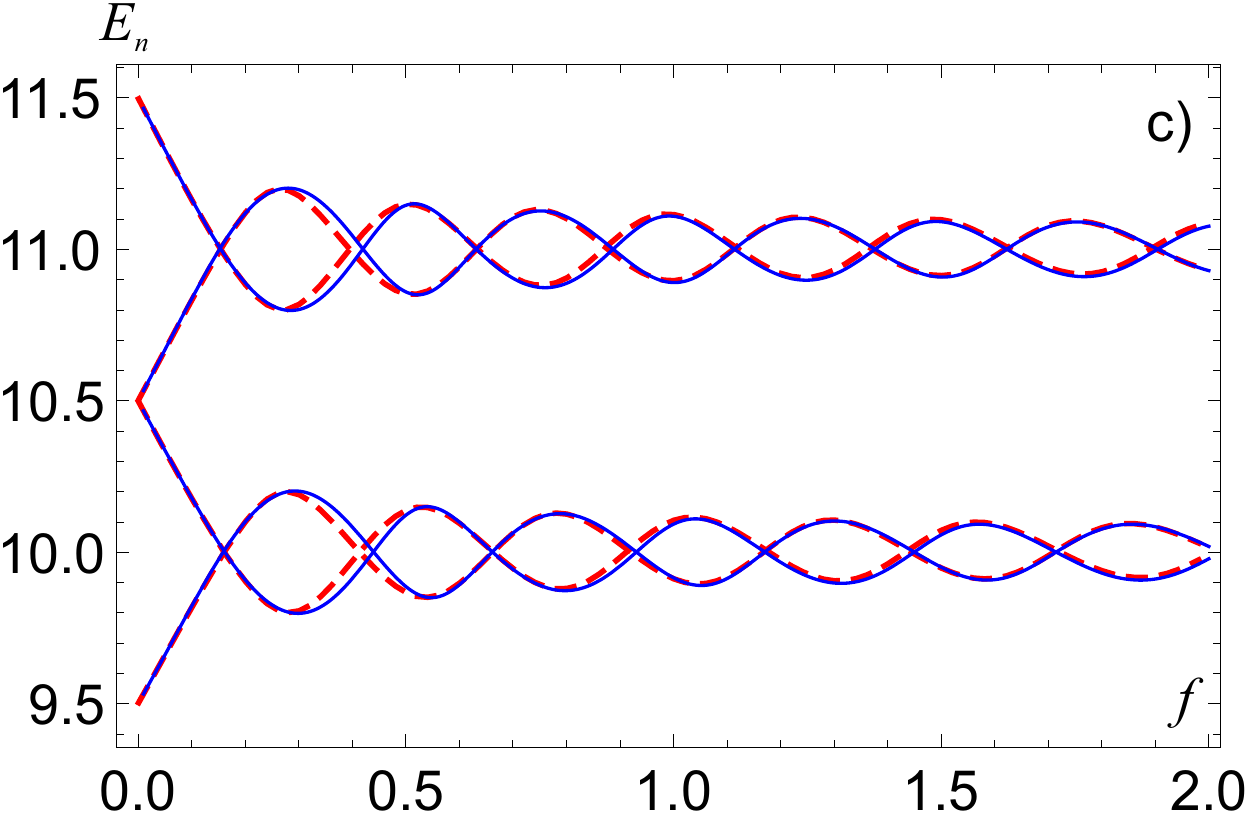} \quad
  \includegraphics[width=.4\linewidth]{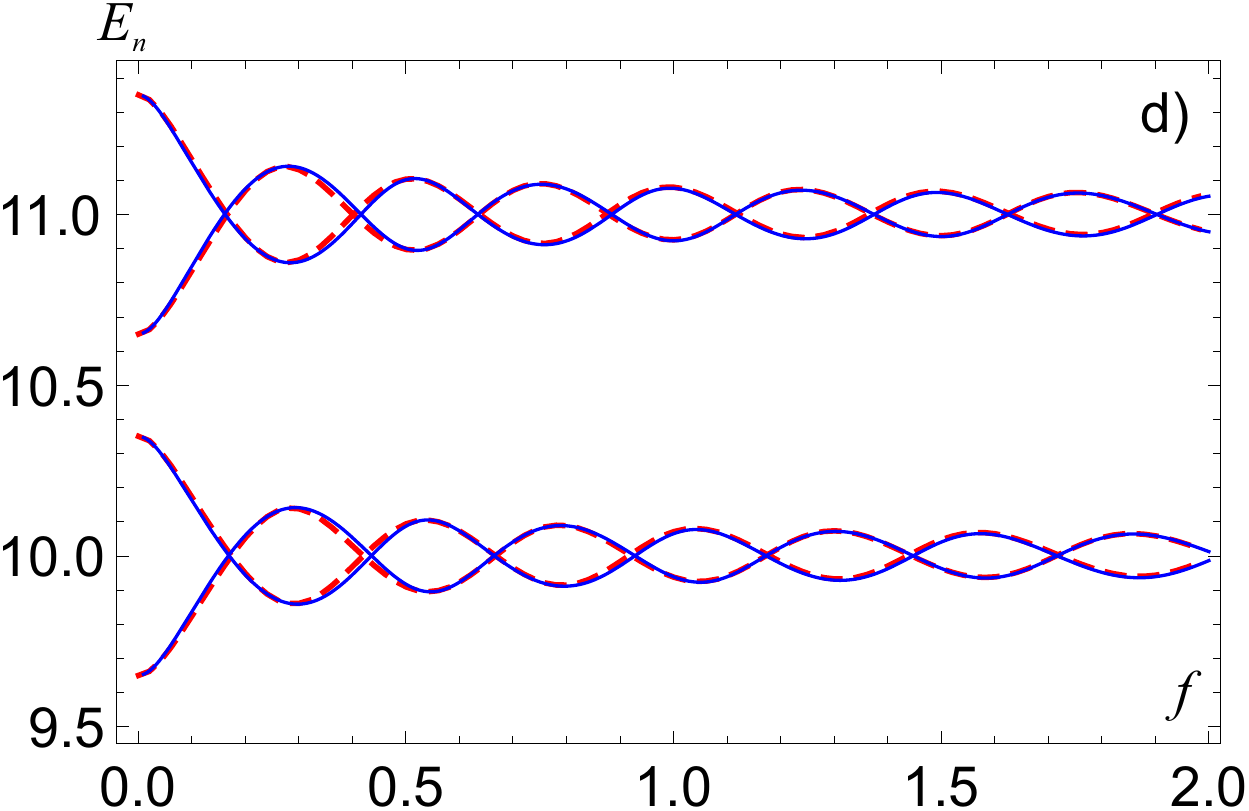}
  \caption{(Color online) Energy levels of QRM as a function of the dimensionless coupling constant $f$ and parameter $\Delta$: (left column) $\Delta = 1.0$, (right column) $\Delta = 0.7$. Low-lying states are shown in (a) and (b), whereas some highly-excited states are presented in (c) and (d). Dashed lines correspond to the numerical solution, solid lines represent the updated zeroth-order approximation of OM (\ref{10}).} 
    \label{fig:2}
\end{figure*}

The coefficients $ A, B $ as well as energy values are determined by the system of linear equations, which can be obtained by substituting the expansion (\ref{9}) into the equation (\ref{6}) and its subsequent projecting onto the vectors $\ket{n} \chi_\uparrow$ and $\ket{n+1} \chi_\downarrow$, respectively (the normalization condition of (\ref{9}) is also implied):
\begin{align}
\label{10}
	E^{(0)}_{n, \pm} = n + \frac{1}{2} +\frac{\Delta}{4}  (-1)^n \bigl( S_{nn}+S_{n+1,n+1} \bigr) \pm \frac{ M}{2},
\end{align}
where
\begin{align}
	M = \Biggl\{ \biggl[ 1 - \frac{\Delta}{2}  (-1)^n \Bigl( S_{nn}-S_{n+1,n+1} \Bigr)  \biggr]^2 \nonumber \\ 
		+ \Delta^2 S_{n,n+1}^2 \Biggr\}^{1/2}; \nonumber \\
	A_{n, \pm} = - \frac{\gamma}{\sqrt{1 + \gamma^2}}; \quad B_{n, \pm} = \frac{1}{\sqrt{1 + \gamma^2}}; \nonumber \\
	\gamma = \frac{\frac{\Delta}{2}(-1)^n S_{n,n+1}}{n+\frac{\Delta}{2}  (-1)^n S_{n,n} - E^{(0)}_{n, \pm}}. \nonumber
\end{align}

In Fig.~\ref{fig:2} we plot the results of calculation by using the formula (\ref{10}), which proves to possess quite high accuracy within the whole range of variation of the QRM parameters. One should also note that formula (\ref{10}) coincides with the results obtained by us earlier \cite{feranchuk_two-level_1996} for the QRM in the dipole gauge by means of using more complicated constructions.

\begin{figure*}[t]
  \includegraphics[width=.4\linewidth]{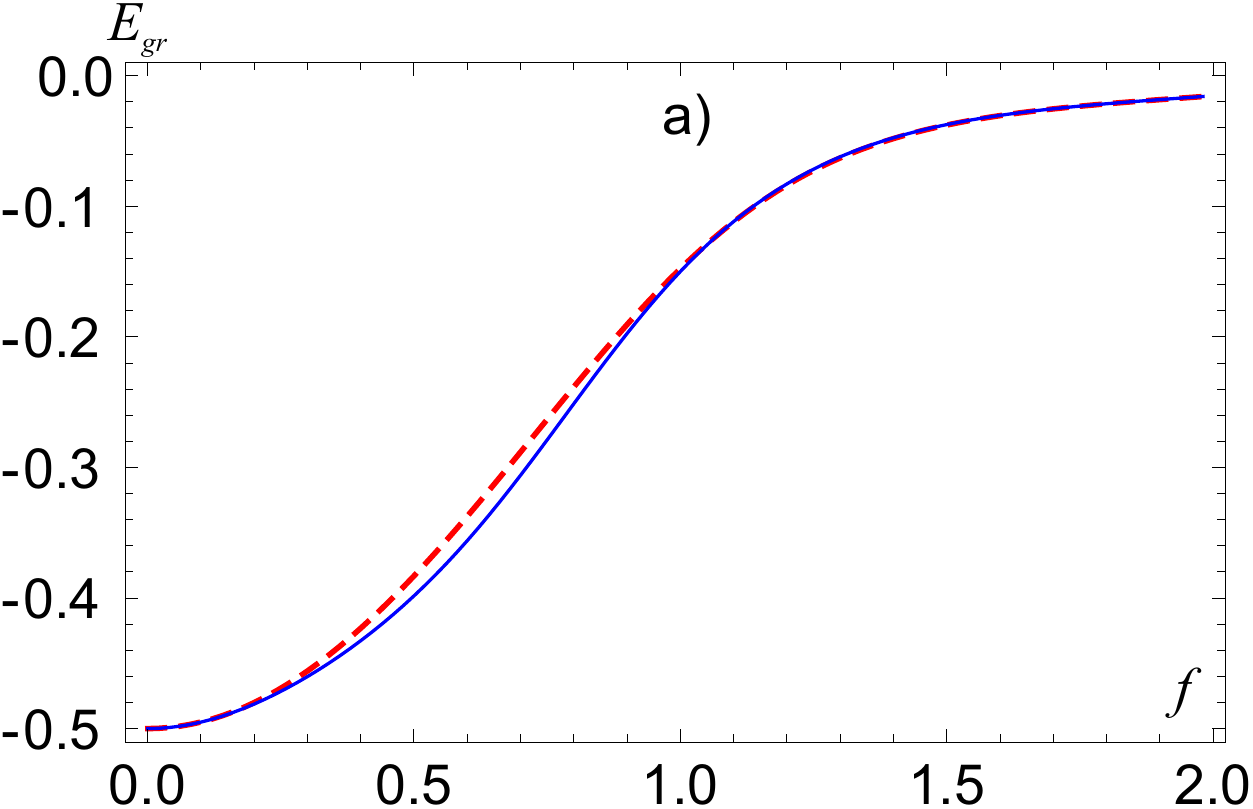} \quad
  \includegraphics[width=.4\linewidth]{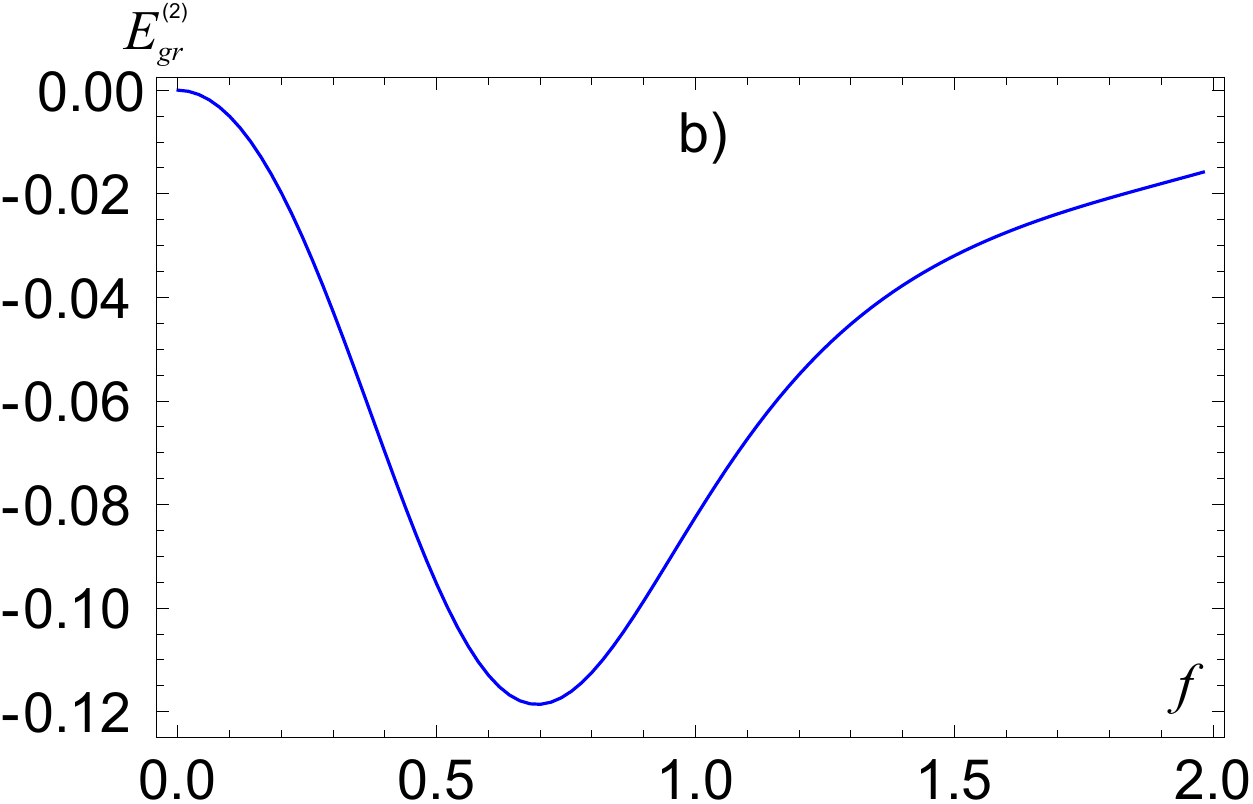} 
  \caption{(Color online) (a) Ground state energy of QRM as a function of the dimensionless coupling constant $f$ and parameter $\Delta = 1$. Dashed lines correspond to the numerical solution, solid lines represent the analytical approximation (\ref{11}). (b) The value of the second-order correction to the ground state energy of QRM as a function of the dimensionless coupling constant $f$ and parameter $\Delta = 1$.}
    \label{fig:3}
\end{figure*}

The simple form of the basis set allows one to perform summation over the intermediate states when calculating the second-order correction to energy levels in order to estimate the accuracy of OM (within OM the first-order correction is identically equal to zero). To calculate the second-order correction to the ground state energy one can effectively use the basis set (\ref{5}): 
\begin{gather}
	E_{\text{gr}} = E_{\text{gr}}^{(0)} + E_{\text{gr}}^{(2)}, \nonumber\\
	E_{\text{gr}}^{(0)} = \braket{\psi_{0,\downarrow}| \hat{H}_C | \psi_{0,\downarrow}} = \braket{ \chi_\downarrow | H_{00} | \chi_\downarrow } \equiv H_{00}^{\downarrow\downarrow}, \nonumber\\
	E_{\text{gr}}^{(2)} = - \sum_{\mu \neq (0,\downarrow)} \frac{|\braket{\psi_\mu |\hat{H}_C| \psi_{0,\downarrow}}|^2}{H_{\mu \mu} - H_{00}^{\downarrow\downarrow}} \nonumber \\
	 = - \sum_{\mu \neq (0,\downarrow)} \frac{|H_{n0}^{s\downarrow}|^2}{H_{kk}^{ss} - H_{00}^{\downarrow\downarrow}}, \quad \mu = (k,s), \label{11}
\end{gather}
whereas addressing the excited states, one should take into account the updated basis set (\ref{9}), which removes the degeneracy of states in the weak coupling range:
\begin{gather}
	E_{nr}= E_{nr}^{(0)} + E_{nr}^{(2)}, \nonumber \\
	E_{nr}^{(2)} = \frac{|F_{nr}|^2}{E^{(0)}_{nr} - E^{(0)}_{n,-r}} + \sum_{\mu \neq \nu_1,\nu_2} \frac{|\widetilde{H}^{sr}_{kn}|^2}{E^{(0)}_{nr} - H^{ss}_{kk}},
\end{gather}
where $\nu_1 = (n,\uparrow)$, $\nu_2 = (n+1,\downarrow)$ and
\begin{align}
\label{12}
	F_{nr} &= \braket{\widetilde{\psi}_{n,-r}| \hat{H}_C | \widetilde{\psi}_{nr}} \nonumber \\
		   &= A_{n,-r}A_{nr} H^{\uparrow\uparrow}_{nn} + A_{n,-r}B_{nr} H^{\uparrow\downarrow}_{n,n+1} \nonumber \\
		   &+ B_{n,-r}A_{nr} H^{\downarrow\uparrow}_{n+1,n} + B_{n,-r}B_{nr} H^{\downarrow\downarrow}_{n+1,n+1}, \nonumber \\
	\widetilde{H}^{sr}_{kn} &= \braket{\psi_\mu| \hat{H}_C | \widetilde{\psi}_{nr}} = A_{nr} H^{s \uparrow}_{kn} + B_{nr} H^{s \downarrow}_{k,n+1}. \nonumber \\
\end{align}

The results of calculation using the formulas (\ref{11}) are shown in Fig.~\ref{fig:3}. As one can see, the obtained approximations practically coincide with the numerical solution.

It should be stressed that in contrast to the standard perturbation theory, the matrix elements in (\ref{11})-(\ref{12}) are calculated with the total Hamiltonian of the system, and not just with the perturbation operator. Using the algebra of creation and annihilation operators, these matrix elements can be calculated analytically using the basis of Fock states.

 \section{DM in the Coulomb gauge for $2$- and $3$-TLS}
 \label{sec:3}
In the paper \cite{di_stefano_resolution_2019} a unitary operator describing the gauge-invariant transformation of  $N$-TLS DM to the Coulomb gauge was also introduced. The corresponding operator of this unitary transformation and the Hamiltonian of $N$-TLS DM have the following explicit form \cite{di_stefano_resolution_2019}:
\begin{gather}
	\hat{U}_N = e^{ 2 f \hat{J}_x (\hat{a}^+ - \hat{a})},
\end{gather}
\begin{multline}
	\hat{H}_N = \hat{a}^+\hat{a} + \Delta \biggl\{ \hat{J}_z \cosh[2f(\hat{a} - \hat{a}^+)] \\ - i \hat{J}_y \sinh[2f(\hat{a}-\hat{a}^+)] \biggr\}, \label{13}
\end{multline}
where  $\hat{J}_y$, $\hat{J}_z $ are the components of the total spin operator of $N$-TLS
\begin{equation}
	\vec{\hat{J}} = \frac{1}{2} \sum_{j=1}^{N} \vec{\hat{\sigma}}^{(j)}.
\end{equation}

\begin{figure*}[t]
  \includegraphics[width=.4\linewidth]{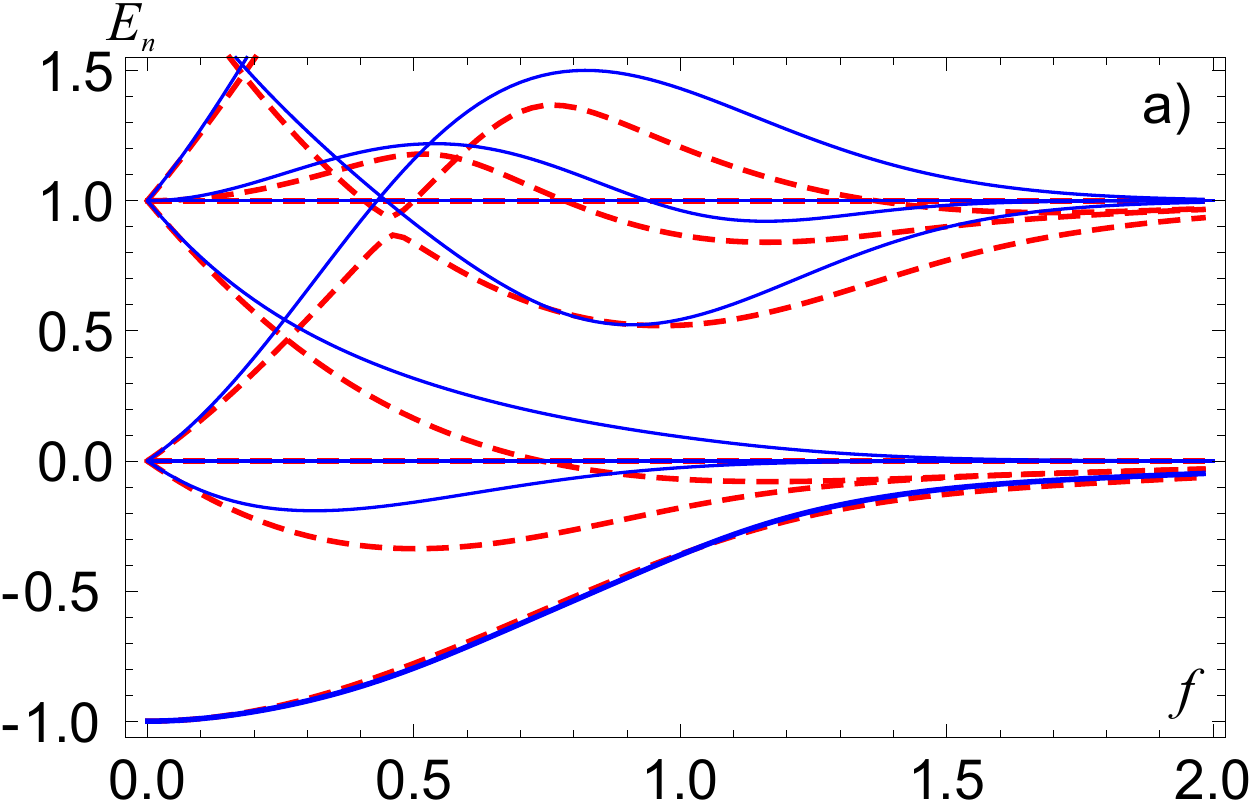} \quad
  \includegraphics[width=.4\linewidth]{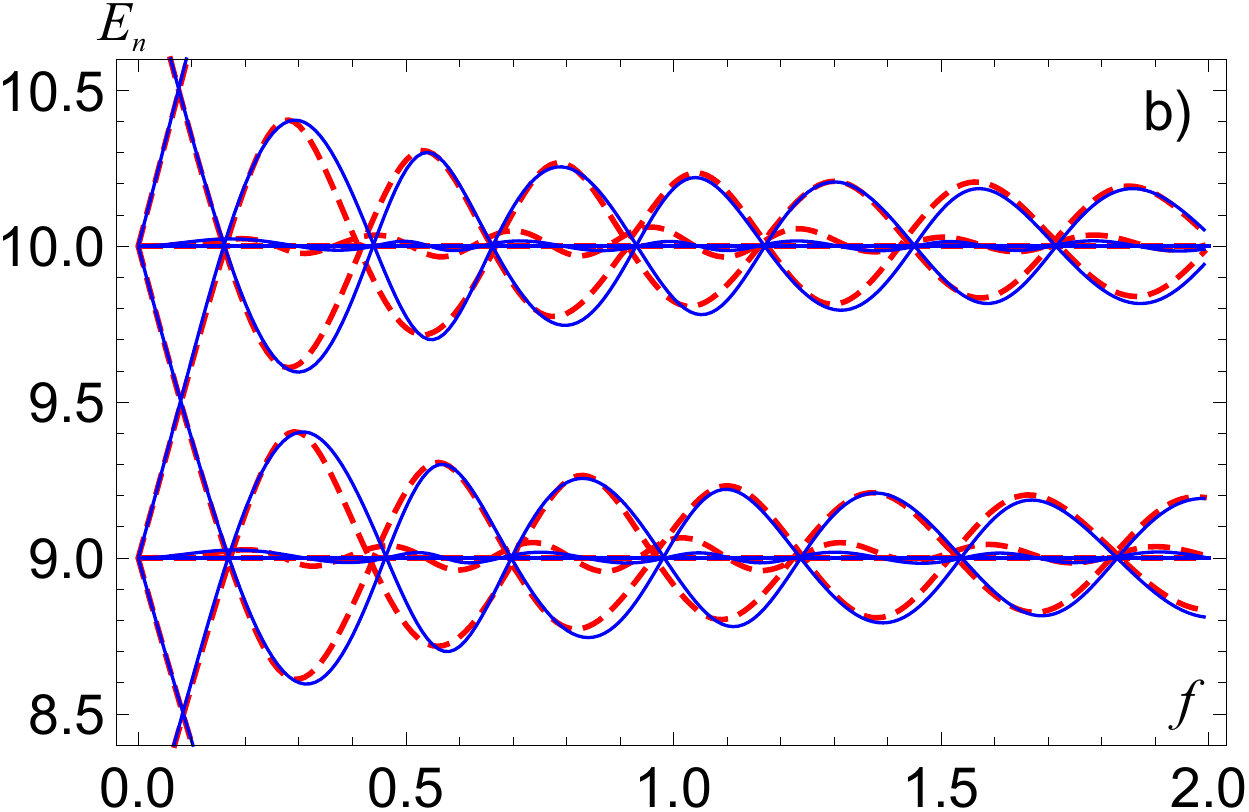} 
  \caption{(Color online) Energy levels of 2-TLS DM as a function of the dimensionless coupling constant $f$ and parameter $\Delta = 1.0$. Low-lying states are shown in (a), whereas some highly-excited states are presented in (b). Dashed lines correspond to the numerical solution, solid lines represent the updated zeroth-order approximation of OM (\ref{17}). The ground state level also includes the second-order correction of OM. Energy levels for both $J=0$ and $J=1$ are shown.} 
    \label{fig:4}
\end{figure*}

Let us first consider a 2-TLS DM within the same approach as for the QRM discussed above. In this case, the total momentum takes the possible values $J = 0$ and $J = 1$, therefore the basis set for deriving the eigenstates of the operator (\ref{13}) consists of the two subsets:
 \begin{gather}
\label{14}
	\ket{\psi}_{n00} = \ket{n}\chi_{00}; \nonumber \\
	\ket{\psi}_{n1M} = \ket{n}\chi_{1M}, \quad M = -1, 0, 1,
\end{gather}
where $\ket{n}$ are the Fock states of the field; $ \chi_{JM} $ are the eigenvectors of operators $\hat{J}^2$, $\hat{J}_z$.

The matrix elements of Hamiltonian (\ref{13}) in the considered basis can be written in the form as follows:
\begin{multline}
	H_{kn} = n \delta_{kn} \text{I}_3 + \frac{\Delta}{2} S_{kn} \Biggl[ \frac{(-1)^n + (-1)^k}{2} J_z \\
	 - i \frac{(-1)^n - (-1)^k}{2} J_y \Biggr], 
\end{multline}
where $\text{I}_3$ is a unit $3 \times 3$ matrix; $J_y$ and $J_z$ are the $3 \times 3$ total spin projection matrices.

The state vectors of the first subset form the exact solutions of the Schrödinger equation for Hamiltonian (\ref{13}) and correspond to the case of non-interacting TLS and the field, so that the energy levels of these states are associated with the field excitations $E_{n00} = n$ only.

Considering the series of states corresponding to the total spin value $J=1$, in the zeroth-order approximation of OM we derive:
 \begin{align}
\label{16}
E^{(0)}_{n1M} &= \braket{\psi_{n1M}|\hat{H}_N|\psi_{n1M}} \nonumber \\
			  &= n + M (-1)^n \Delta S_{nn} (f).
\end{align}

As in the case of QRM, the accuracy of the zeroth-order approximation can be substantially improved by using the correct linear combinations of states with different projections of the momentum and the same energy in the absence of interaction. Similarly to (\ref{9}), such modified state vectors have the following form
\begin{gather}
	\ket{\psi_{\text{gr}}} = \ket{0} \chi_{1,-1}; \nonumber \\
	\ket{\widetilde{\psi}_{1p}} =  A_{1p} \ket{\psi_{110}}+ B_{1p} \ket{\psi_{011}}, \ p = 1,2; \nonumber\\
	\ket{\widetilde{\psi}_{np}} =  A_{np} \ket{\psi_{n1,-1}}+ B_{np} \ket{\psi_{n-1,10}} \nonumber \\
	 + C_{np} \ket{\psi_{n-2,11}}, \  p = 1,2,3, \,  n \geq 2.
	 \label{17}
\end{gather}

\begin{figure*}[t]
  \includegraphics[width=.4\linewidth]{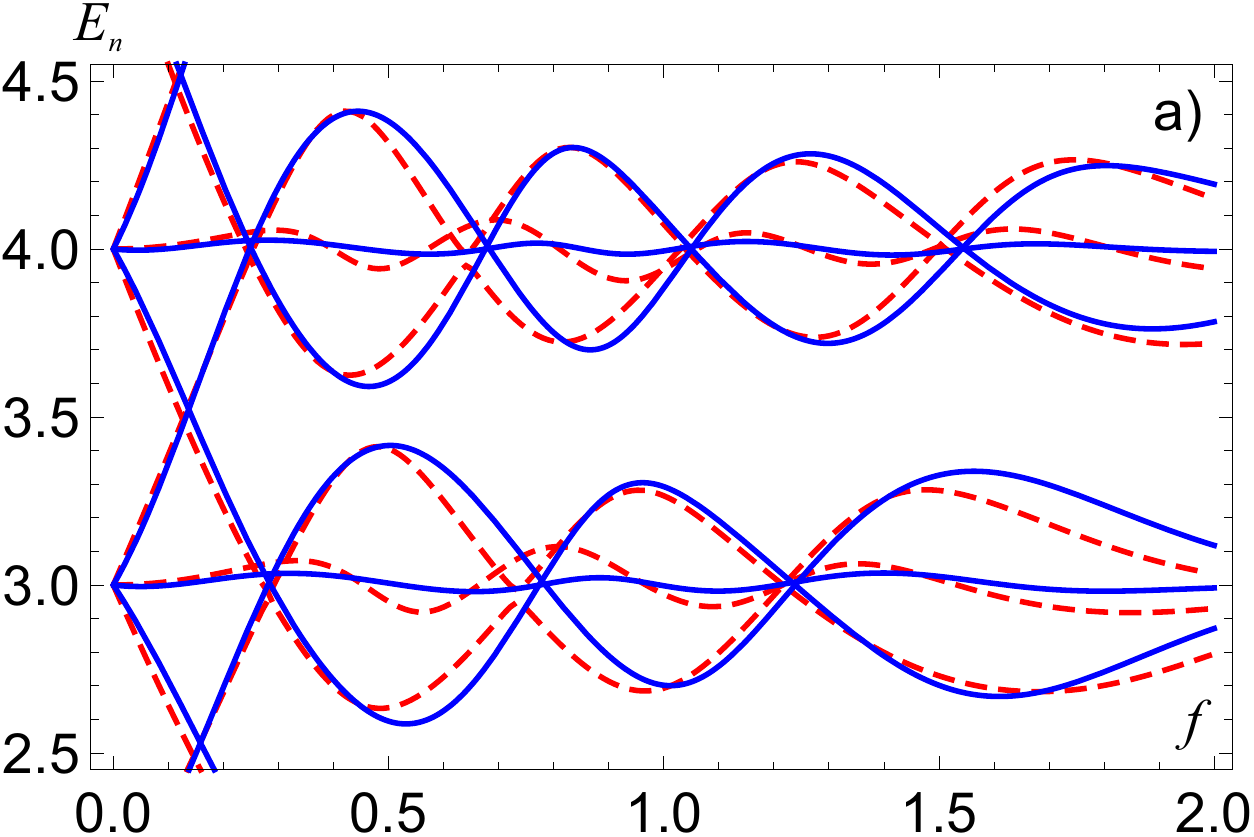} \quad
  \includegraphics[width=.4\linewidth]{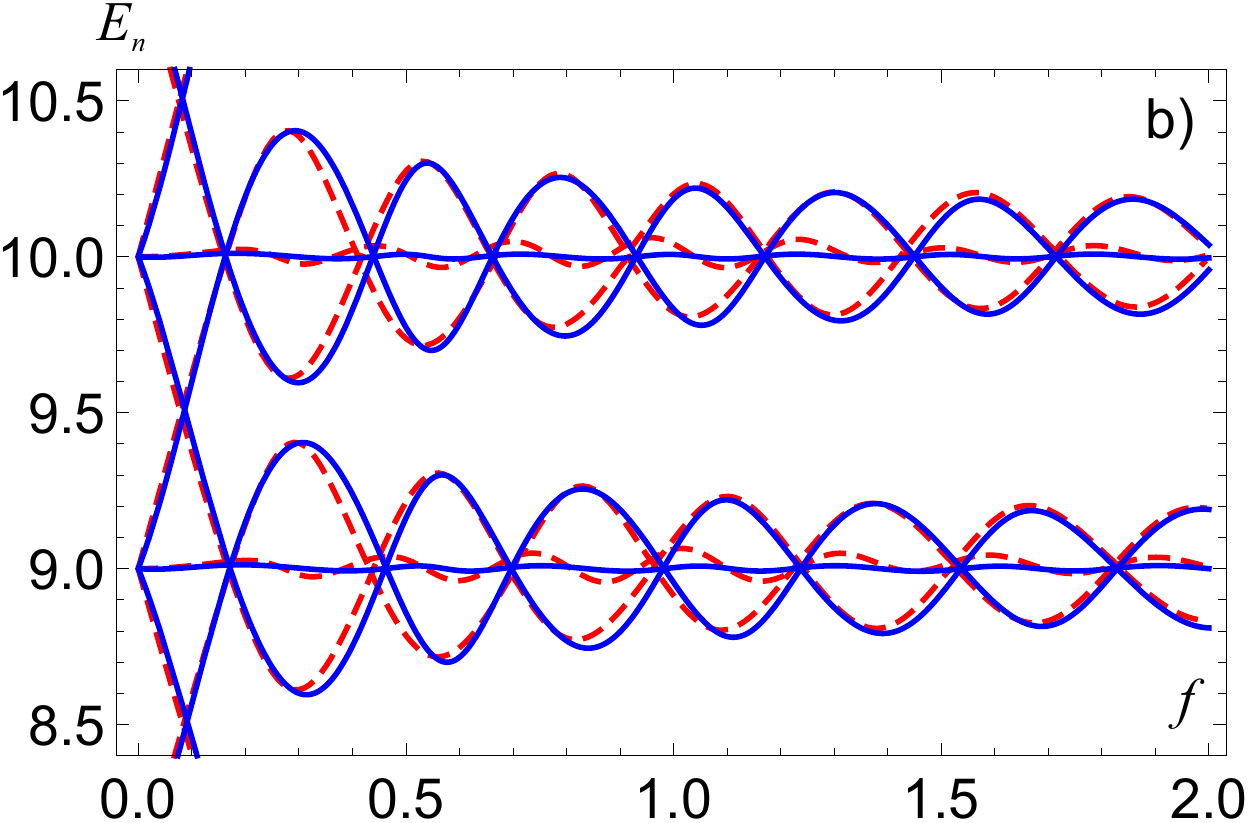} 
  \caption{(Color online) Energy levels of 2-TLS DM as a function of the dimensionless coupling constant $f$ and parameter $\Delta = 1.0$. Low-lying states are shown in (a), whereas some highly-excited states are presented in (b). Dashed lines correspond to the numerical solution, solid lines represent the truncated zeroth-order approximation of OM (\ref{17b}). Energy levels for $J=1$ are shown.} 
    \label{fig:5}
\end{figure*}

The coefficients $A_{1p}$, $B_{1p}$ can be defined in a similar way as it was done in the case of QRM resulting in expression (\ref{10}), whereas the coefficients $A_{np}$, $B_{np}$, and $C_{np}$ are defined as the components of normalized eigenvectors of the following matrix:
\begin{eqnarray}
\label{18}
	\hat F_2 =
	\begin{pmatrix}
		H_{nn}^{-1,-1} & H_{n,n-1}^{-1,0} & H_{n,n-2}^{-1,1} \\
		(H_{n,n-1}^{-1,0})^* & H_{n-1,n-1}^{00} & H_{n-1,n_2}^{01} \\
		(H_{n,n-2}^{-1,1})^* & (H_{n-1,n_2}^{01})^* & H_{n-2,n-2}^{11}
	\end{pmatrix},
\end{eqnarray}
where $H_{kn}^{MM'} = \braket{\chi_{1M}|H_{kn}|\chi_{1M'}}$.

In Fig.~\ref{fig:4} we plot the energy spectrum of the 2-TLS DM calculated on the basis of (\ref{17})-(\ref{18}) and compare it with the results of numerical calculation. As one can conlude, the analytical approximation (\ref{17}) reproduces the main features of the numerical solution, and the accuracy improves for the highly excited states. We also include the second-order correction for the ground state, so that the analytical solution almost coincides with the numerical one.

In the considered case the eigenvalues of (\ref{18}) are calculated by solving the cubic equation and their analytical form is bulky. However, it appears to be possible to find the simple approximation for them if the following linear combinations of the degenerate states are used instead of $\ket{\widetilde{\psi}_{np}}$:
\begin{align}
\label{17a}
	\ket{\overline{\psi}_{n1}} &=  A_{n1} \ket{\psi_{n,1,-1}} + B_{n1} \ket{\psi_{n-1,10}},  \nonumber\\
	\ket{\overline{\psi}_{n2}} &=  B_{n2} \ket{\psi_{n-1,1,0}} + C_{n2} \ket{\psi_{n-2,11}}.
\end{align}

In this case  the roots of two quadratic equations define the  analytical approximations for 3 eigenvalues in the way as follows:
\begin{gather}
	\overline{E}_{n1}= E^+_{n1}; \ \overline{E}_{n2} = \frac{1}{2}(E^-_{n1} + E^+_{n2}); \ \overline{E}_{n3} = E^-_{n2} ;   \nonumber\\
	E^{\pm}_{n1} = \frac{1}{2} \biggl( H_{nn}^{-1,-1} + H_{n-1,n-1}^{00} \nonumber \\
	\pm \sqrt{(H_{nn}^{-1,-1} - H_{n-1,n-1}^{00})^2 + 4 |H_{n,n-1}^{-1,0}|^2} \biggr); \nonumber\\
	E^{\pm}_{n2} = \frac{1}{2} \biggl( H_{n-1,n-1}^{00} + H_{n-2,n-2}^{11} \nonumber \\
	\pm \sqrt{(H_{n-1,n-1}^{00} - H_{n-2,n-2}^{11})^2 + 4 |H_{n-1,n-2}^{01}|^2} \biggr).
\label{17b}
\end{gather}

In Fig.~\ref{fig:5} we plot the results of approximation (\ref{17b}), which prove to interpolate the numerical values fairly well.

The 3-TLS DM can be considered in a similar way. In this case, the total momentum takes the possible values $J= 1/2$ and $J = 3/2$. Therefore, the basis set for finding the eigenstates of operator (\ref{13}) consists of two subsets of the following form:
\begin{gather}
	\ket{\psi_{n,\frac{1}{2},M}} = \ket{n}\chi_{\frac{1}{2},M}, \,\, M=\pm 1/2; \nonumber \\
	\ket{\psi_{n,\frac{3}{2},M}} = \ket{n}\chi_{\frac{3}{2},M}, \,\, M = \pm 1/2, \pm 3/2,
\label{19}
\end{gather}
where the same notation as in (\ref{17}) is used.

It can be demonstrated both analytically and numerically that the series of levels corresponding to the first subset coincides with the energy levels of QRM.

Let us consider the series of states corresponding to $J = 3/2$. In this case, the correct linear combinations in the zeroth-order approximation of OM have the form as follows:
\begin{gather}
	\ket{\psi_{\text{gr}}} = \ket{0}\chi_{\frac{3}{2},- \frac{3}{2}}; \nonumber\\
	\ket{\widetilde{\psi}_{1p}} =  A_{1p} \ket{\psi_{1,\frac{3}{2},-\frac{3}{2}}}+ B_{1p} \ket{\psi_{0,\frac{3}{2},-\frac{1}{2}}}, \,\, p = 1,2; \nonumber\\
	\ket{\widetilde{\psi}_{2p}} =  A_{2p} \ket{\psi_{2,\frac{3}{2},-\frac{3}{2}}}+ B_{2p} \ket{\psi_{1,\frac{3}{2},-\frac{1}{2}}} \nonumber\\
	  + C_{2p} \ket{\psi_{0,\frac{3}{2},\frac{1}{2}}}, \,\,  p = 1,2,3; \nonumber\\
	\ket{\widetilde{\psi}_{np}} =  A_{np} \ket{\psi_{n,\frac{3}{2},-\frac{3}{2}}}+ B_{np} \ket{\psi_{n-1,\frac{3}{2},-\frac{1}{2}}} \nonumber\\
	  + C_{np} \ket{\psi_{n-2,\frac{3}{2},\frac{1}{2}}}+ D_{np} \ket{\psi_{n-3,\frac{3}{2},\frac{3}{2}}}, \nonumber\\ \,\, p = 1,2,3,4, \,\,  n \geq 3.
\label{20}
\end{gather}

The coefficients $A_{1p}$, $B_{1p}$ can be defined in a similar way as it was done in the case of QRM resulting in expression (\ref{10}), and the coefficients $A_{np}$, $B_{np}$, and $C_{np}$ can be derived similarly to 2-TLS DM and expression (\ref{17}). The main series of coefficients $A_{np}$, $B_{np}$, $C_{np}$, $D_{np}$ are defined as the components of the eigenvectors of the $ (4 \times 4) $ matrix with a composition similar to (\ref{18}), which we do not present here for shortness.

\begin{figure*}[t]
  \includegraphics[width=.4\linewidth]{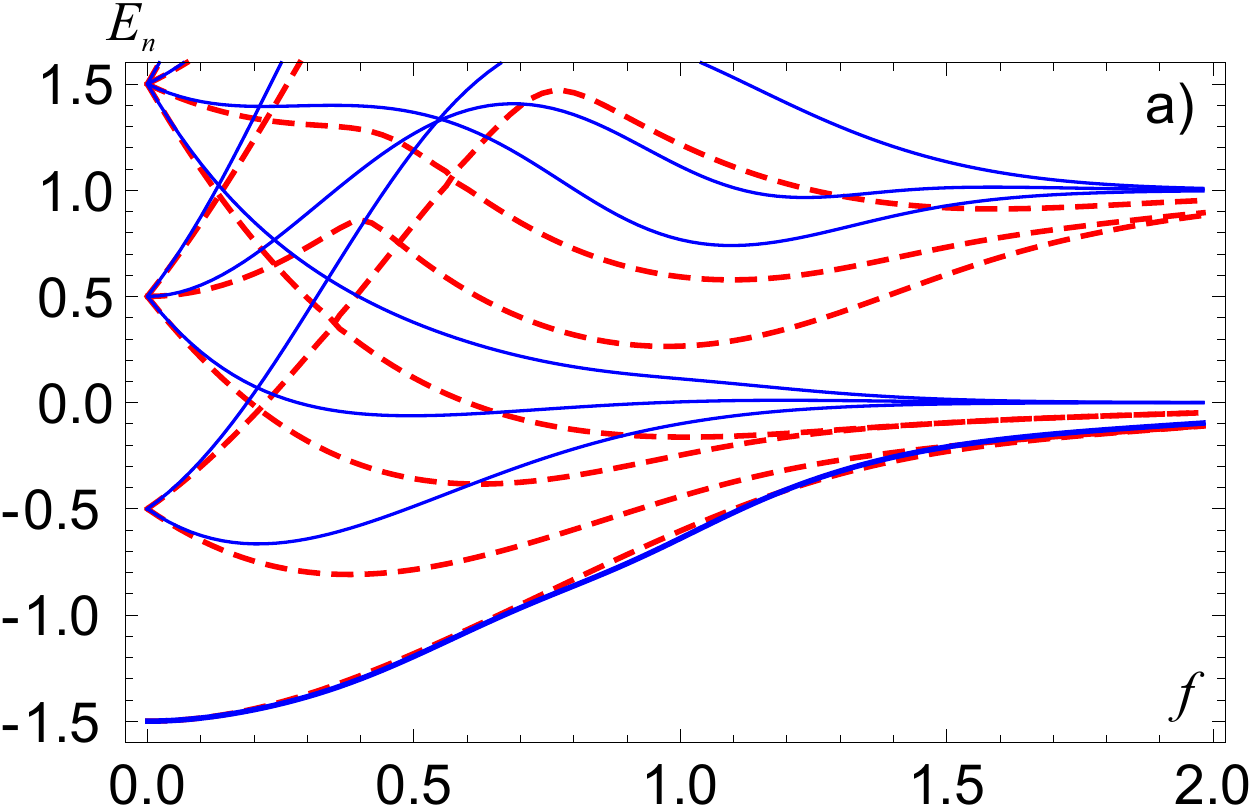} \quad
  \includegraphics[width=.4\linewidth]{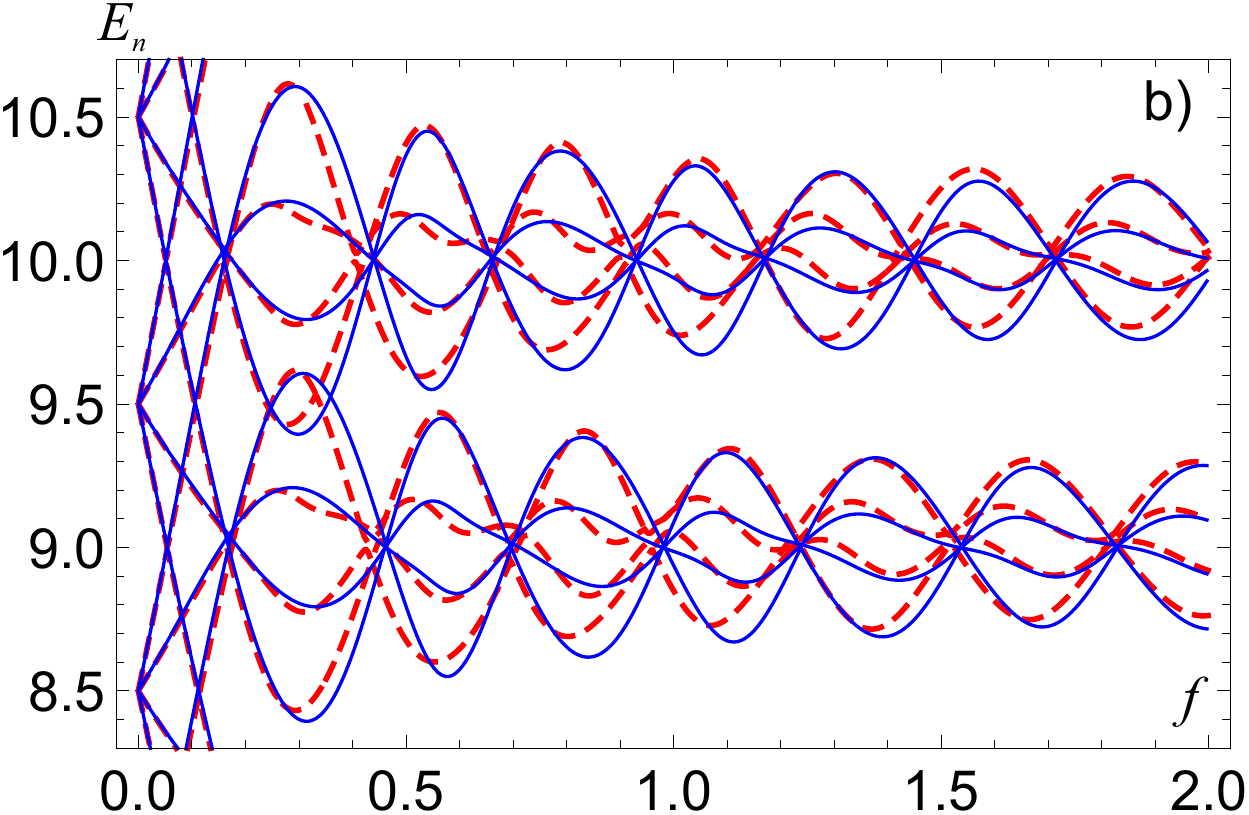} \break\break
  \includegraphics[width=.4\linewidth]{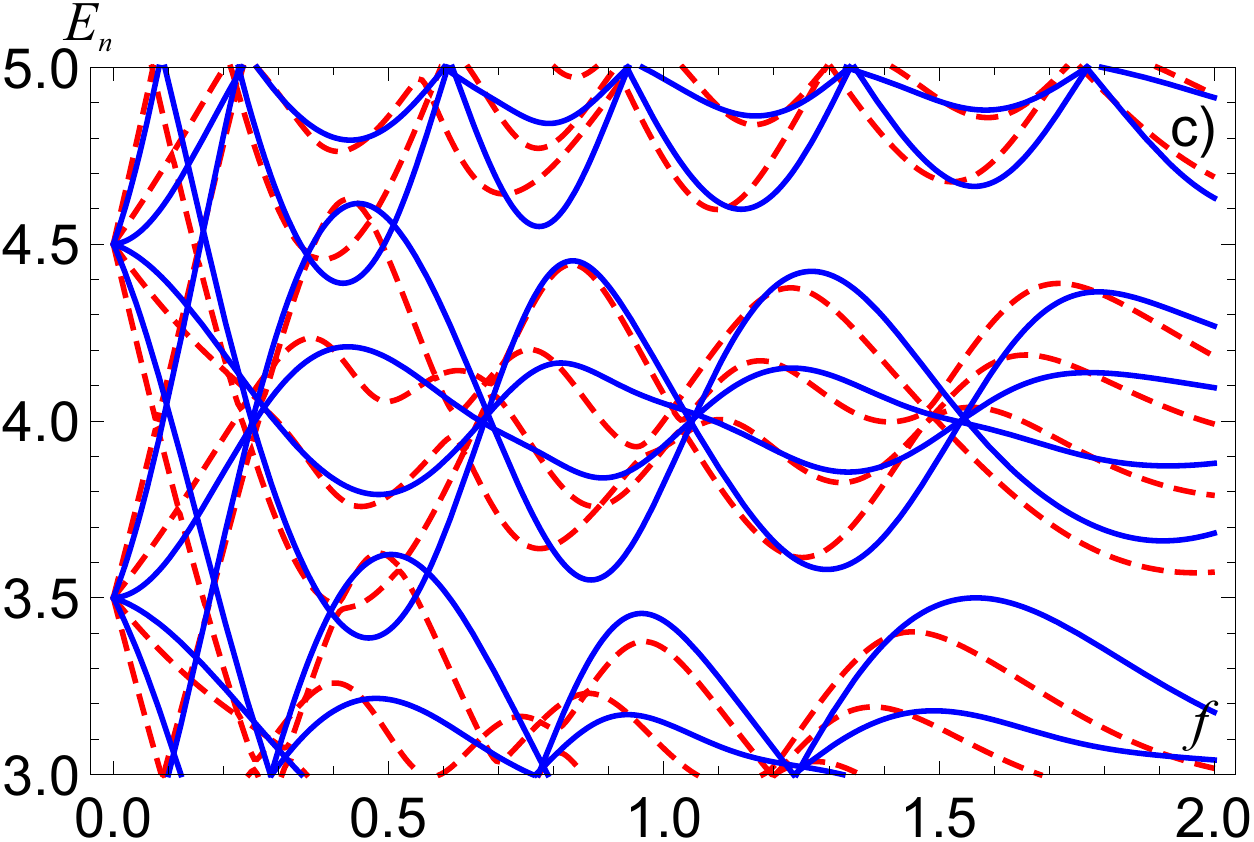} \quad
  \includegraphics[width=.4\linewidth]{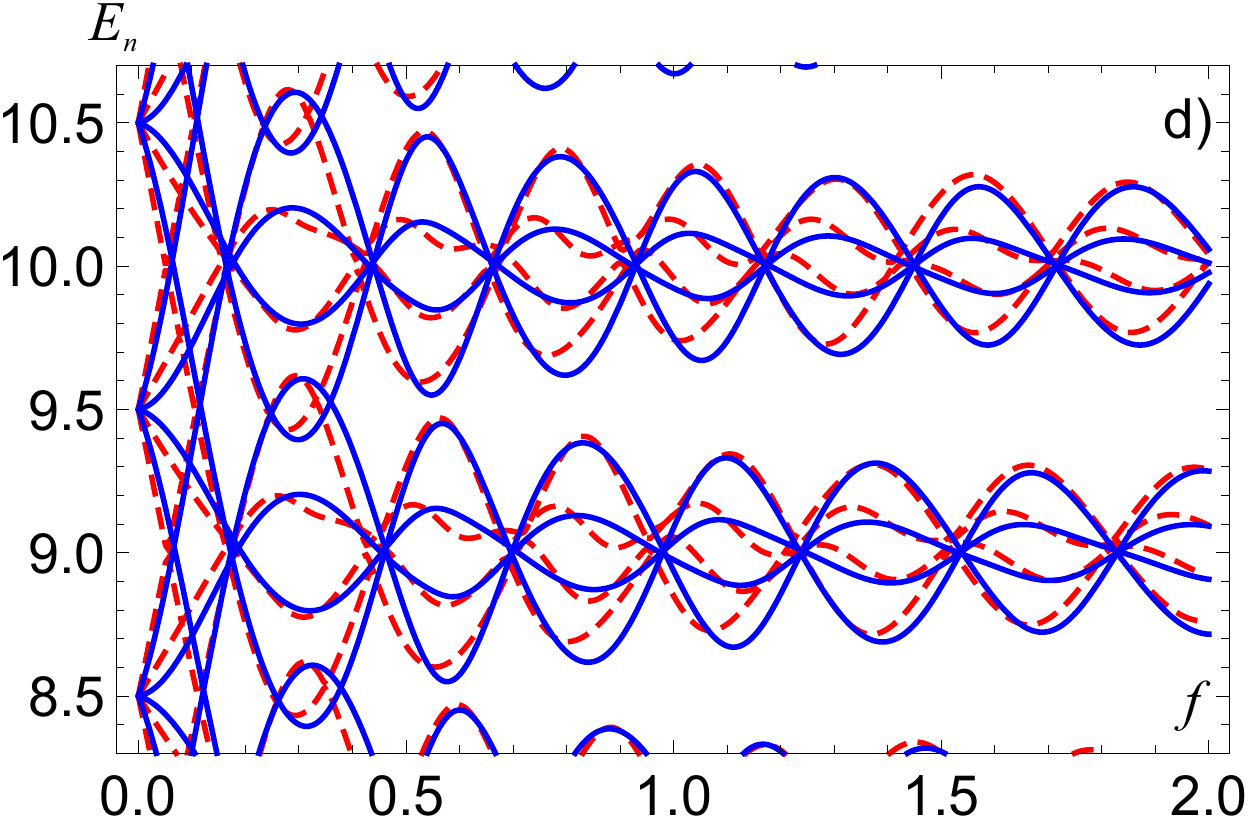}
  \caption{(Color online) Energy levels of 3-TLS DM as a function of the dimensionless coupling constant $f$ and parameter $\Delta = 1.0$.  Low-lying states are shown in (a) and (c), whereas some highly-excited states are presented in (b) and (d). Dashed lines correspond to the numerical solution, solid lines in (a) and (b) represent the updated zeroth-order approximation of OM (\ref{20}), solid lines in (c) and (d) represent the truncated zeroth-order approximation of OM (\ref{21}). The ground state level in (a) also includes the second-order correction of OM. Energy levels for $J=3/2$  are shown.}
  \label{fig:6}
\end{figure*}

Moreover, in the case of 3-TLS DM it is still possible to split the wave vector $\ket{\widetilde{\psi}_{np}}$ into the 3 linear combinations, each including only 2 basis vectors. It leads to the analytical approximation for the eigenvalues in the following form:
\begin{gather}
	\overline{E}_{n1}= E^+_{n1}; \,\, \overline{E}_{n2} = \frac{1}{2}(E^-_{n1} + E^+_{n2}); \nonumber\\
	\overline{E}_{n3} = \frac{1}{2}(E^-_{n2} + E^+_{n3}); \,\, \overline{E}_{n4} = E^-_{n3} ;  \label{21}
\end{gather}
\begin{multline*}
	E^{\pm}_{n1} = \frac{1}{2} \Biggl( H_{nn}^{-\frac{3}{2},-\frac{3}{2}} + H_{n-1,n-1}^{-\frac{1}{2},-\frac{1}{2}} \\
		\pm \sqrt{(H_{nn}^{-\frac{3}{2},-\frac{3}{2}} - H_{n-1,n-1}^{-\frac{1}{2},-\frac{1}{2}})^2 + 4 |H_{n,n-1}^{-\frac{3}{2},-\frac{1}{2}}|^2} \Biggr);
\end{multline*}
\begin{multline*}
	E^{\pm}_{n2} = \frac{1}{2} \Biggl( H_{n-1,n-1}^{-\frac{1}{2},-\frac{1}{2}} + H_{n-2,n-2}^{\frac{1}{2},\frac{1}{2}}\\
		\pm \sqrt{(H_{n-1,n-1}^{-\frac{1}{2},-\frac{1}{2}} - H_{n-2,n-2}^{\frac{1}{2},\frac{1}{2}})^2 + 4 |H_{n-1,n-2}^{-\frac{1}{2},\frac{1}{2}}|^2} \Biggr);
\end{multline*}
 \begin{multline}
	E^{\pm}_{n3} = \frac{1}{2} \Biggl(  H_{n-2,n-2}^{\frac{1}{2},\frac{1}{2}} + H_{n-3,n-3}^{\frac{3}{2},\frac{3}{2}} \nonumber\\
		\pm \sqrt{(H_{n-2,n-2}^{\frac{1}{2},\frac{1}{2}} - H_{n-3,n-3}^{\frac{3}{2},\frac{3}{2}})^2 + 4 |H_{n-2,n-3}^{\frac{1}{2},\frac{3}{2}}|^2} \Biggr).
\end{multline}

\begin{figure*}[tbh]
  \includegraphics[width=.4\linewidth]{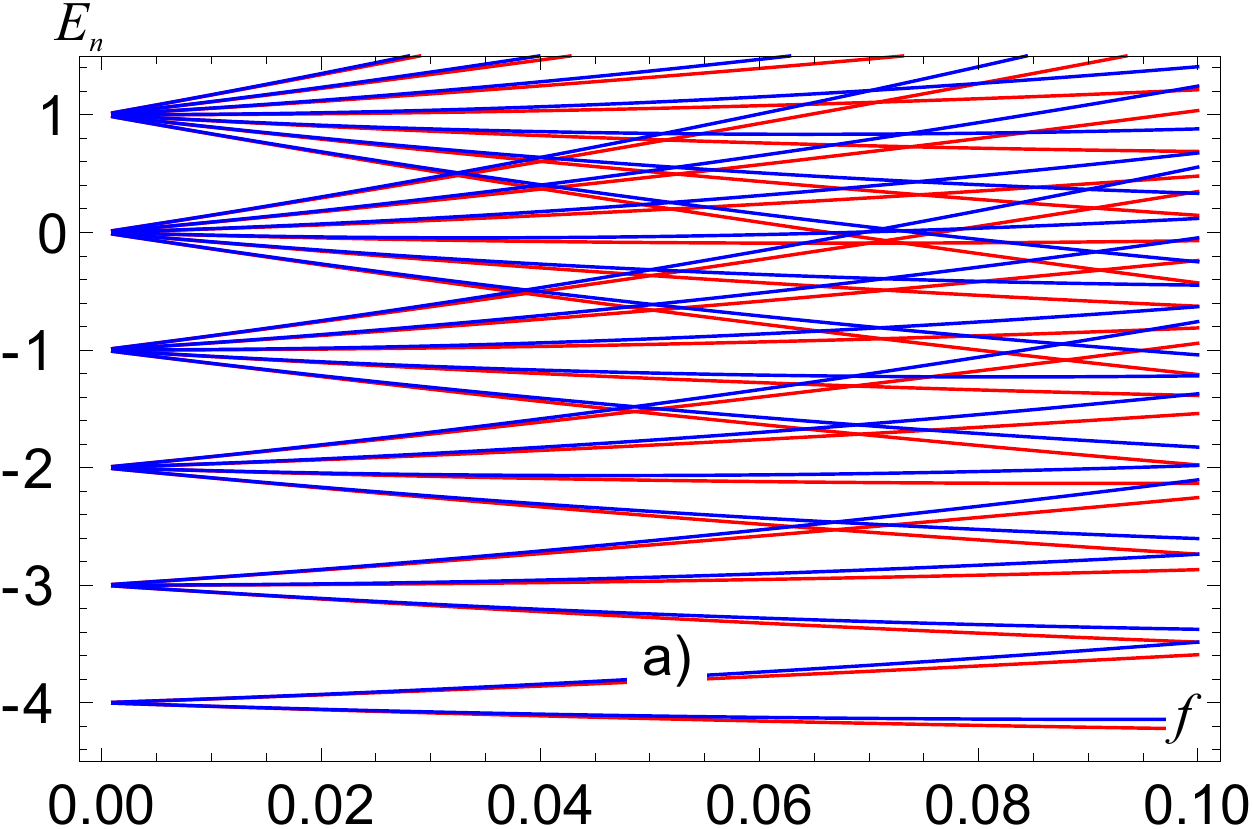} \quad
  \includegraphics[width=.4\linewidth]{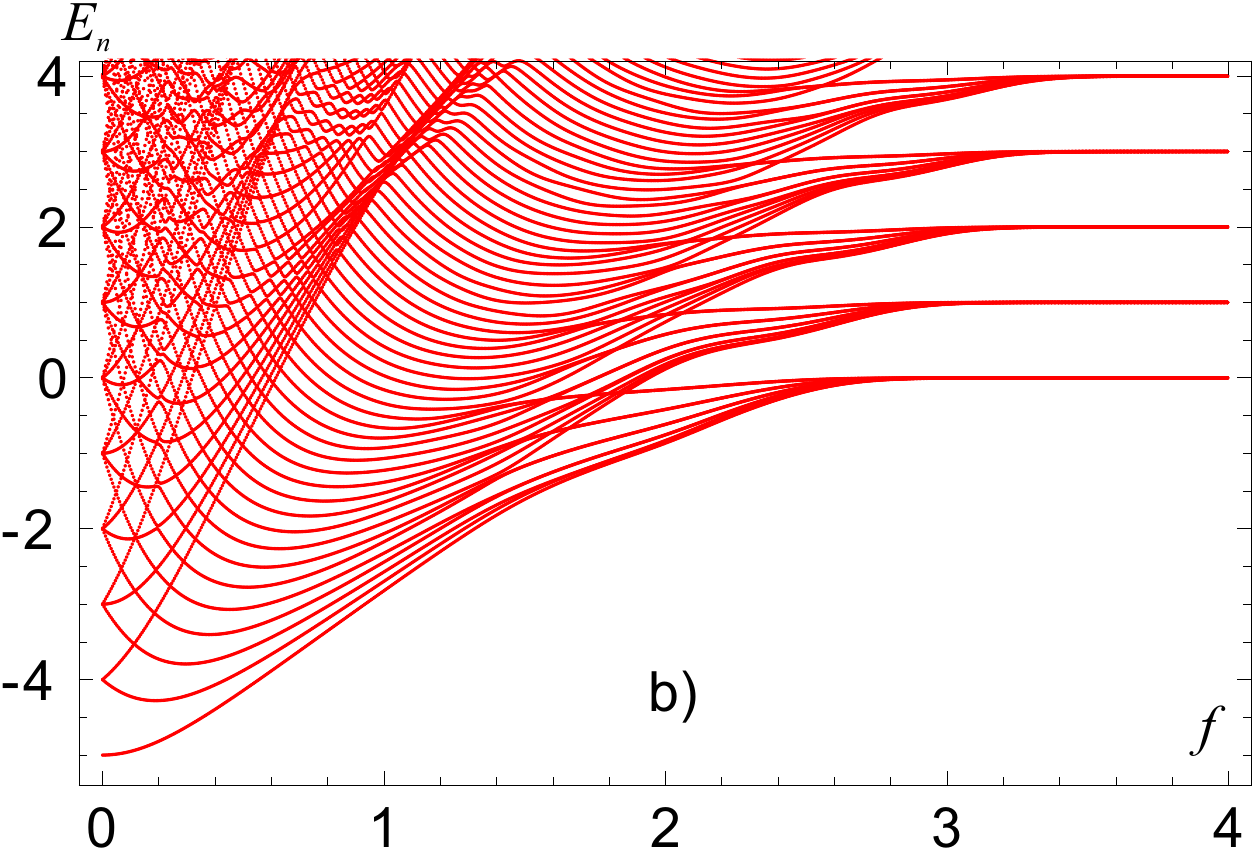} 
  \caption{(Color online) Energy levels of 10-TLS DM as a function of the dimensionless coupling constant $f$ and parameter $\Delta = 1.0$: (a) weak coupling range; (b) wide range including the ultra-strong copuling regime. Red lines correspond to the numerical solution, blue lines represent the updated zeroth-order approximation of OM. Energy levels for $J=5$ are shown.} 
    \label{fig:7}
\end{figure*}

In Fig.~{\ref{fig:6}} we show the results of calculating the energy levels of the main series. One can see that the approximations based on the wave vectors $\ket{\widetilde{\psi}_{np}}$ and on the analytical expressions (\ref{21}) have almost the same accuracy. Analytical solution restores the qualitative behavior of the numerical solution for the low-lying states, whereas the accuracy increases when considering the highly-excited states. Additional feature of this case is that the levels corresponding to the projections $ M = \pm 1/2 $ are doubly degenerate for any coupling constant. This is due to the two options of obtaining these states during summation of 3 spin moments.

\section{Dicke model in the Coulomb gauge for $N$-TLS}
 \label{sec:4}

The algorithm of constructing the approximation for the eigenstates considered above can be generalized for DM with arbitrary number of TLS. However, it should be stressed that for the large amount of atoms, such an approximation becomes ineffective for real applications. The fact is that the calculation of correct linear combinations, similar to (\ref{20}), requires the diagonalization of matrices of dimension $ (N + 1) \times (N + 1) $ or solutions of $(N+1)$ quadratic equations. In both cases manipulating these approximations requires numerical calculations. Taking into account the up-to-date algorithms of numerical diagonalization of high-dimensional matrices, it becomes more efficient to directly diagonalize the complete Hamiltonian in the matrix representation using the basis set in the form of a product of one-particle vectors of type (\ref{5}). The characteristic form of the matrix elements is determined by the expressions (\ref{7}). The accuracy can be determined by the condition that the obtained eigenvalues should not vary with the further increase of the size of the matrix. Diagonalization of the Hamiltonian Matrix was carried out using the Wolfram Mathematica software and took only several miliseconds for a $ (220 \times 220)$ matrix.

As an example, in Fig.~\ref{fig:7} we show the energy levels of the 10-TLS DM for the main series corresponding to the total momentum of $J = 5$. We have considered separately the range of the weak coupling $f \leq 0.1$ where the analytical approximations are still quite effective and the wide range of $f$ including the deep strong coupling where the eigenvalues are defined by the asymptotical expression $E_{ns} \approx n$.

It should also be stressed that energy spectrum of DM with $N \gg 1$ acquires the universal character, which can be appoximately defined as the following scaling transformation
\begin{eqnarray}
\label{22}
	E (N,n,f) \approx N E (1,nN,f).
\end{eqnarray}

Fig.~\ref{fig:8} shows the comparison of the energy spectra for $E(10,2n,f)$ and $E(20,n,f)/2$. One can conclude that these spectra mostly satisfy the relation (\ref{22}) within the plane of variables $(n,f)$. This property could be essential for analyzing the properties of the system in the thermodynamical limit.
\begin{figure*}[tbh]
  \includegraphics[width=.6\linewidth]{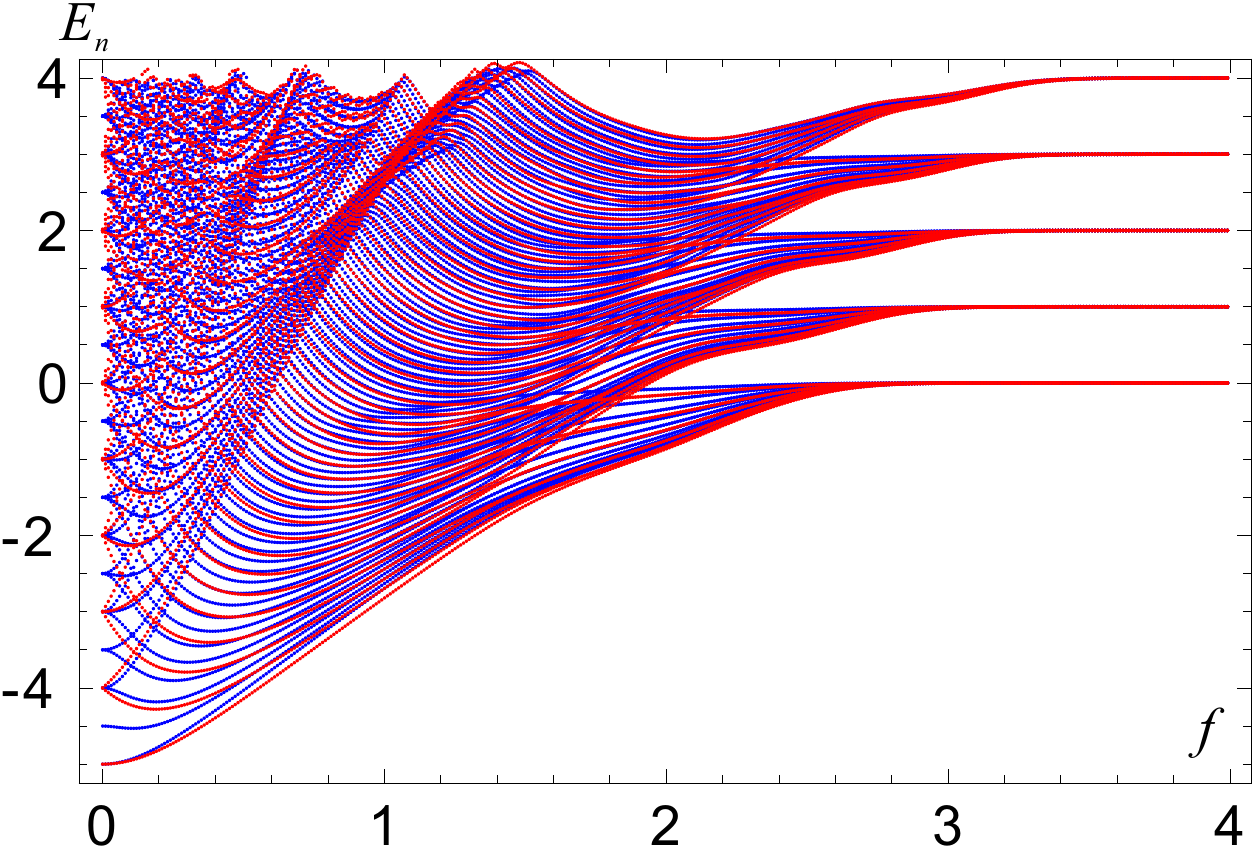} 
  \caption{(Color online) Scaled energy levels of $N$-TLS DM satisfying the relation (\ref{22}) as a function of the dimensionless coupling constant $f$ and parameter $\Delta = 1.0$. Red lines correspond to the 10-TLS DM and $E_n = E(10,2n,f)$; blue lines represent 20-TLS DM and $E_n = E(20,n,f)/2$.} 
    \label{fig:8}
\end{figure*}

\section{Conclusions and Outlook}

In our work we have analyzed the spectrum of arrays of two-level systems which interact with a single-mode quantum field in a cavity in the wide range of the coupling constant values. It is shown that the complete set of simple basis vectors allows one to find both the analytical approximation and the effective algorithm for the numerical calculation of the system's eigenvalues.

The obtained results can be useful for the analysis of evolution of the atomic polarization and some other characteristics as field fluctuations, entanglement of the qubits, etc. As a next step, we are going to consider the thermodynamics of the Dicke model in order to check whether the gauge invariant Hamiltonian does not lead to the phase transition in the system.

\section{Acknowledgements}  

Authors are grateful to Professor Salvatore Savasta for the discussion about the parameters of the gauge invariant Hamitonian. We also acknowledge the Belarusian State Program of Scientific Researches for the financial support.    

\end{large}

\section{References}
\label{sec:ref}






\bibliography{biblqrmdm}

\end{document}